\documentclass[prb,twocolumn,showpacs,preprintnumbers]{revtex4}
\usepackage{graphicx}
\usepackage{subfigure}
\usepackage{amsfonts}
\usepackage{amssymb}
\usepackage{amsmath}
\usepackage{bm}

\begin{document}

\title{Resonant state expansion applied to planar open optical systems}

\author{ M.\,B. Doost}
\author{W. Langbein}
\author{E.\,A. Muljarov}
\altaffiliation[]{egor.muljarov@astro.cf.ac.uk; on leave from General Physics Institute RAS,
Moscow, Russia} \affiliation{School of Physics and Astronomy, Cardiff University, Cardiff CF24 3AA,
United Kingdom}
\begin{abstract}

The resonant state expansion (RSE), a novel perturbation theory of
Brillouin-Wigner type developed in electrodynamics
[Muljarov, Langbein, and Zimmermann, Europhys. Lett., {\bf 92}, 50010
(2010)], is applied to planar, effectively one-dimensional optical systems, such as layered
dielectric slabs and Bragg reflector microcavities.  It is demonstrated that the RSE converges with
a power law in the basis size. Algorithms for error estimation and their reduction by extrapolation
are presented and evaluated. Complex eigenfrequencies, electro-magnetic fields, and the Green's
function of a selection of optical systems are calculated, as well as the observable
transmission spectra. In particular we find that for a Bragg-mirror microcavity, which has sharp
resonances in the spectrum, the transmission calculated using the resonant state expansion
reproduces the result of the transfer/scattering matrix method.
\end{abstract}


\date{\today}

\maketitle

\section{Introduction}

Recently, a novel perturbation method for the treatment of open electromagnetic systems, the
resonant states expansion (RSE), has been formulated.\cite{Muljarov10} Unlike previous
perturbative approaches\cite{Lai90,Lai91,Leung94,Leung96,Yang10} which due to their complexity and poor convergency
are limited to small perturbations, this method is shown to be suitable for perturbations of
arbitrary strength and shape. It is based on the concept of resonant states (RS) of an open system,
also known in quantum mechanics as Gamow\cite{Gamow28} or Siegert
states.\cite{Siegert39,Moiseyev98} These states exponentially decay in time and grow in space at
large distances.\cite{Baz69} Owing to their completeness inside a finite area of space, RS can be
used for expanding solutions of the Maxwell equations, reducing the wave problem for the modes of
the system to diagonalization of a finite complex matrix. Hence the strength and the shape of the
perturbed dielectric profile that can be treated is limited only by the size of the chosen finite
basis of unperturbed RS.

The idea of resonances is a cornerstone of physics, allowing to rationalize the dynamic behavior
of physical systems. In open systems, excitations decay with time, endowing resonances with a
spectral width. Depending on their width and separation resonances appear in measured spectra as isolated
lines or merge into a continuum. The concept of RS provides a unified picture of an open system
which includes all types of resonances and is an alternative to the commonplace division of the
spectrum into non-decaying bound and continuum states with real energies. RS are discrete
eigenstates which have complex frequencies (and equivalently energies and wave numbers) and satisfy
outgoing wave boundary conditions. This corresponds to a physical situation that an open system,
excited at an earlier time, looses its energy to the outside space. The imaginary part of the
frequency reflects the temporal decay of the energy in the system. Owing to this leakage, the RS
wave functions have tails (outgoing waves) which grow exponentially outside of the system and
cannot be normalized by the usual integration of their square modulus. Instead, the normalization
and orthogonality of RS is given by an integral over the finite volume of the system and the energy
flux to the outside in the form of a surface term.\cite{Siegert39,Weinstein69}

The presence of a continuum in the spectrum of a system is a significant problem for any
perturbation theory. In open electromagnetic systems such a continuum is often the dominating if
not the only part of the spectrum. However, going away from the real axis to the complex frequency
plane, the continuum can in many cases be effectively replaced by a countable number of discrete RS
which form a complete basis.
Therefore, RS of a perturbed system can be expanded into the unperturbed RS. The expansion
coefficients can be found by diagonalizing a complex symmetric matrix which consists of a diagonal
matrix representing the bare spectrum, and the perturbation.\cite{Muljarov10} The perturbed
resonant states can then be used to calculate the Green's function of the system via its spectral
representation,\cite{Newton60,More71} using the Mittag-Leffler theorem. The Green's function
provides the complete system response and allows to calculate observables such as emission,
scattering, or transmission. This recently formulated general method of reducing the Maxwell
equations for an open optical system to a linear matrix eigenvalue problem is called resonant state
expansion (RSE).

The RSE has been suggested\cite{Muljarov10} as an appropriate tool for calculation of sharp
resonances in optical spectra, such as perturbed whispering gallery modes of a dielectric
microsphere. Popular computational techniques in electrodynamics, such as the finite difference in
time domain (FDTD)\cite{Taflove00,Hagness97} or the finite element
method\cite{Wiersig03,Zienkiewicz00,Rahman91} adapted to such problems, require a large
computational domain in time and/or space, and can produce spurious solutions.\cite{Jiang96} In
particular, sharp resonances are characterized by optical modes which decay slowly in time and
hence FDTD needs a large time domain. Furthermore, being applied to open systems, the finite
element method either introduces a significant error when the boundary is too close, or needs to
consider an excessively large domain in real space in order to describe the far-field asymptotics
correctly. The RSE does not suffer from these problems because it produces the eigenstates of the
system, and in particular their wave numbers, directly by diagonalization of a matrix determined
by the near-field properties only.

In this paper, we apply the RSE method outlined in Section~\ref{sec:method} to various planar optical systems. For such effectively one-dimensional (1D) systems efficient alternative methods exist to calculate the transmission and
reflection, enabling verification of the RSE results. We investigate the accuracy with which
eigenfrequencies, eigenfunctions, the Green's function, and transmission can be reproduced. We give
a method to evaluate the convergence and to extrapolate the results in Section~\ref{sec-convergence}. We
apply the RSE to a perturbed dielectric slab in Section~\ref{sec:results} with two different kind of
perturbations: a wide-layer perturbation, and a $\delta$-perturbation. We find that the RSE converges to the exact
solution with a power law in the basis size. As an example of a structure with a sharp resonance,
we treat a Bragg mirror microcavity in Section~\ref{subsec:microcavity}.

\section{The Perturbation Method}\label{sec:method}
The system of Maxwell's equations for a planar dielectric structure with permeability $\mu=1$ surrounded by
vacuum is reduced to the following wave equation:
\begin{equation}
\partial^{2}_{z}{\mathbb E}_{\nu}(z,t)=\bigl[\varepsilon(z)+\Delta\varepsilon(z)\bigr]{\partial^{2}_{t}{\mathbb E}_{\nu}(z,t)}\,,
\label{MXW}
\end{equation}
where $\varepsilon(z)$ denotes the unperturbed dielectric profile, and $\Delta\varepsilon(z)$ the
perturbation of the dielectric function.  Here the transverse eigenmodes with index $\nu$ are taken
with zero in-plane wave number. The electric field ${\mathbb E}_{\nu}(z,t)$  can be written in an
harmonic form
\begin{equation}
{\mathbb E}_{\nu}(z,t)=\mathcal{E}_{\nu}(z)\exp(-ic\varkappa_{\nu}t)
\end{equation}
with complex frequency $c\varkappa_{\nu}$ ($c$ is the speed of light in vacuum) and amplitude
$\mathcal{E}_{\nu}(z)$ satisfying the following time-independent wave equation:
\begin{equation}
\Bigl\{\partial^{2}_{z}+\bigl[\varepsilon(z)+\Delta\varepsilon(z)\bigr]\varkappa_{\nu}^2\Bigr\}\mathcal{E}_{\nu}(z)=0\,.
\label{pslab}
\end{equation}
The electric field $\mathcal{E}_{\nu}(z)$ and its first derivative are continuous everywhere. RS are
eigenmodes which satisfy outgoing boundary conditions, given by the form
\begin{equation}\label{BC}
\mathcal{E}_{\nu}(z)=A^\pm_\nu \exp(i\varkappa_\nu|z|)
\end{equation}
in the surrounding vacuum with the amplitudes $A^-_\nu$ (on the left-hand side) and  $A^+_\nu$ (on
the right-hand side of the structure) which are generally different. In the case of a
mirror-symmetric system, $A^-_\nu=A^+_\nu$  for symmetric and $A^-_\nu=-A^+_\nu$ for antisymmetric
modes.

In the following, for the unperturbed RS, {\em i.e.} for $\Delta\varepsilon(z)=0$,
$\mathcal{E}_{\nu}(z)$ is denoted as $E_{n}(z)$, and $\varkappa_{\nu}$ as $k_n$. The unperturbed RS
are orthogonal and normalized according to
\begin{eqnarray}
&&\int_{-a}^{a} \varepsilon(z){E}_{n} (z){E}_{m} (z)\,dz \nonumber\\
&&- \frac{{E}_{n} (-a){E}_{m} (-a) +{E}_{n} (a){E}_{m} (a)}{i({k}_{n}+{k}_{m})}=\delta_{nm}\,,
\label{nint}
\end{eqnarray}
where $z=\pm a$ are the positions of the boundaries of the unperturbed system. The perturbed states
are written as linear combinations of the normalized unperturbed RS,
\begin{equation}
\mathcal{E}_{\nu}(z)=\sum_{n}c_{n\nu}\,\frac{E_n(z)}{\sqrt{k_n}}\,, \label{superp}
\end{equation}
resulting in the linear eigenvalue problem for $\varkappa_{\nu}$ and $c_{n\nu}$
\begin{equation}
\sum_{m}\Big(\frac{\delta_{nm}}{k_n}+\frac{V_{nm}}{2\sqrt{k_nk_m}}\Big)c_{m\nu}=\frac{1}{\varkappa_{\nu}}c_{n\nu}
\label{diag}
\end{equation}
with the perturbation matrix\cite{Muljarov10}
\begin{equation}
V_{nm}=\int_{-a}^{a}{\Delta\varepsilon(z)E_n(z)E_m(z)\,dz}\,. \label{pmatrix}
\end{equation}

In a dielectric system with real refractive index, the RS wave numbers $k_n$ have the following
general property: Im\,$k_n\leq0$ and ${\rm Re}\,k_{-n}={\rm Re}\,k_{+n}$. Additionally, in 1D systems
(planar systems at normal incidence) there is always a RS with ${\rm Re}\,k_0=0$ and
Im\,$k_0\neq0$. We number the RS with increasing real part of their wave number, numbering the
state with zero real part as state number zero. The number of RS in the unperturbed or perturbed
systems is countable infinite. Therefore we always deal will a
truncation of the basis of the RS, which is the only approximation of the theory. We refer to
$n_{\rm{max}}$ as the truncation number for the basis so that $-n_{\rm{max}} \leq n\leq
n_{\rm{max}}$. Hence the basis size $N$ is given by
\begin{equation}
\label{nmax} N=2{n_{\rm{max}}}+1\,.
\end{equation}
Ideally,
by choosing the basis size $N$ sufficiently large, the results of the perturbation theory can be
produced with any given accuracy.

The unperturbed system can be any convenient system. In the discussed 1D case, a dielectric slab in
vacuum having thickness $2a$ and real dielectric constant
\begin{equation}
\varepsilon(z)=\left\{
\begin{array}{cl}
\epsilon_s &\text{for } \left|z\right|< a\\
1 & \text{otherwise }
\end{array} \right.\label{unpslab}
\end{equation}
is the simplest system having an analytic solution. We use it as unperturbed system in the following.
The expressions for the unperturbed RS are given in the Appendix. The dielectric constant is taken
to be $\sqrt{\epsilon_s}=1.5$ unless otherwise stated.

\section{Convergence and extrapolation}
\label{sec-convergence}

We introduce a method to estimate the convergence and to extrapolate the RS wave numbers calculated
via the perturbation theory $\varkappa^{(N)}_{\nu}$ to their exact values
$\varkappa_{\nu}^{\rm{(exact)}}$. To do so, we approximate the absolute error in each wave number
as a power law in the basis size $N$:
\begin{equation}
\varkappa_{\nu}^{\rm{(exact)}}-\varkappa_\nu^{(N)}\approx K'_\nu N^{\alpha'_\nu}\approx
K''_\nu N^{\alpha''_\nu}. \label{errorone}
\end{equation}
We assume that the exponent in the power law ($\alpha'_\nu$ or $\alpha''_\nu$) is a real
number, so that the RS wave numbers converge in a straight line in the complex plane.  To determine
the coefficients and exponents of the two representations in  Eq.\,(\ref{errorone}) we use four different values of $N$:
$N_{1}<N_{2}<N_{3}<N_{4}$, where
\begin{equation}
N_1\approx \eta^4 {N_4}\,,\quad N_2\approx \eta^2{N_4}\,,\quad N_3\approx \eta{N_4}\,,
\end{equation}
producing four sets of wave numbers, with the factor $0<\eta<1$.  We match states between the four sets sequentially, {\em
i.e.} first $\{ \varkappa^{(N_4)}_{\nu} \}$ to $\{ \varkappa^{(N_3)}_{\nu} \}$, then $\{
\varkappa^{(N_3)}_{\nu} \}$ to $\{ \varkappa^{(N_2)}_{\nu} \}$, and finally $\{ \varkappa^{(N_2)}_{\nu} \}$
to $\{ \varkappa^{(N_1)}_{\nu} \}$. In doing this, we use the following matching algorithm (MA)
between two sets of wave numbers, $\{ \varkappa^{\rm{(A)}}_{\nu} \}$  and $\{
\varkappa^{\rm{(B)}}_{\nu} \}$:
\begin{itemize}
\item[(a)] Determine the distance between the complex wave numbers of all pairs with one element from $\{ \varkappa^{\rm{(A)}}_{\nu} \}$ and one element from $\{ \varkappa^{\rm{(B)}}_{\nu} \}$.

\item[(b)] Select the pair with the shortest distance, store it, and remove it from the sets.

\item[(c)] Repeat (b) until $\{ \varkappa^{\rm{(A)}}_{\nu} \}$ or $\{ \varkappa^{\rm{(B)}}_{\nu} \}$ is empty.
\end{itemize}
This procedure results in $N_1$ vectors
$(\varkappa^{(N_1)}_{\nu},\varkappa^{(N_2)}_{\nu},\varkappa^{(N_3)}_{\nu},\varkappa^{(N_4)}_{\nu})$
of RS wave numbers. The specific factors chosen between $N_1$, $N_2$, $N_3$, and $N_4$ allow for the following
analytical expressions for two sets of coefficients and exponents in Eq.\,(\ref{errorone}), for each
state $\nu$:
\begin{eqnarray}
\alpha'_\nu&=&\frac{1}{2\ln\eta}\ln\left(\left|\frac{{\varkappa^{(N_4)}_{\nu}}-{\varkappa^{(N_1)}_{\nu}}}{{\varkappa^{(N_4)}_{\nu}}-{\varkappa^{(N_2)}_{\nu}}}\right|-1\right),
\label{errorsix}
\\
\alpha''_\nu&=&\frac{1}{\ln\eta}\ln\left(\left|\frac{{\varkappa^{(N_4)}_{\nu}}-{\varkappa^{(N_2)}_{\nu}}}{{\varkappa^{(N_4)}_{\nu}}-{\varkappa^{(N_3)}_{\nu}}}\right|-1\right),
\label{errorsix1}
\\
K'_\nu&=&\frac{{\varkappa^{(N_4)}_{\nu}}-{\varkappa^{(N_2)}_{\nu}}}{{N}^{\alpha'_{\nu}}_2-{N}^{\alpha'_{\nu}}_4}\,,
\label{errorseven}
\\
K''_\nu&=&\frac{{\varkappa^{(N_4)}_{\nu}}-{\varkappa^{(N_3)}_{\nu}}}{{N}^{\alpha''_{\nu}}_3-{N}^{\alpha''_{\nu}}_4}\,.
\label{errorseven1}
\end{eqnarray}
For extrapolation of eigenvalues and estimation of errors 
we also introduce mean values $\alpha_\nu$ and $K_\nu$ defined as
\begin{equation}
\alpha_\nu=\frac{\alpha'_\nu+\alpha''_\nu}{2}\,,\ \ \ K_\nu N^{\alpha_\nu}_4=\frac{K'_\nu N^{\alpha'_\nu}_4+K''_\nu N^{\alpha''_\nu}_4}{2}\,.
\end{equation}

In order to test the quality of our power law fit, we estimate for each state $\nu$ the relative
extrapolation error defined as
\begin{equation}
F_\nu=\Phi(K'_\nu N^{\alpha'_\nu}_4,K''_\nu N^{\alpha''_\nu}_4)\,, \label{Quality}
\end{equation}
where $2\Phi(X,Y)=\Gamma(X,Y)+\Gamma(Y,X)$ and
\begin{equation}
\Gamma(X,Y)=\left|\frac{X}{Y}-1\right|. \label{Gamma}
\end{equation}
Indeed, $F_\nu$ has the meaning of a relative error in the power law approximation of the distance
$\varkappa^{\rm{(exact)}}_{\nu}-{\varkappa^{(N_4)}_{\nu}}$ deduced from the two sets of power law
parameters. If this error is sufficiently small, $F_{\nu} < F_{\rm{max}}$, and the power law
converges ($\alpha_\nu<\alpha_{\rm max}$), we can improve the result calculated for the
largest basis size $N_4$ by extrapolating it towards the exact value,
$\varkappa^{(N_4)}_{\nu}\to\varkappa^{(\infty)}_{\nu}$, where the extrapolated wave vector
$\varkappa^{(\infty)}_{\nu}$ is defined according to Eq.\,(\ref{errorone}) as
\begin{equation}
\varkappa^{(\infty)}_{\nu}=\varkappa^{(N_4)}_{\nu}+K_{\nu}N^{\alpha_{\nu}}_4.
\label{Improved}
\end{equation}
Otherwise, the power law is not describing the convergence well. We then use the absolute variation
scaled to the system size to evaluate
 if the state has sufficiently converged
\begin{equation}
M_{\nu}= \max_{i=1,2,3}\left|{\varkappa^{(N_4)}_{\nu}}-{\varkappa^{(N_i)}_{\nu}}\right|a\,.
\end{equation}

We use state $\nu$ for the calculation of the Greens function if its relative or absolute error is
sufficiently small, {\em i.e.} if one of the two selection criteria (SC) is met:
\begin{enumerate}
 \item\label{Accept1}{extrapolation error $F_\nu|K_{\nu}N_4^{\alpha_{\nu}}|a<M_{\rm{max}}$ provided that $F_{\nu}< F_{\rm{max}}$ and $\alpha_{\nu}< \alpha_{\rm max}$}\,;
 \item\label{Accept2}{absolute error $M_{\nu}<M_{\rm{max}}$}\,.
\end{enumerate}
For the results shown in the present paper we used $M_{\max}=0.1$, $F_{\rm{max}}=1$, $\alpha_{\max}=-0.5$,
and $\eta=2^{-1/4}$. 
\section{Results}\label{sec:results}

\subsection{Wide-layer perturbation}
\label{widelayer}

\begin{figure}[t]
\includegraphics*[width=0.9\columnwidth]{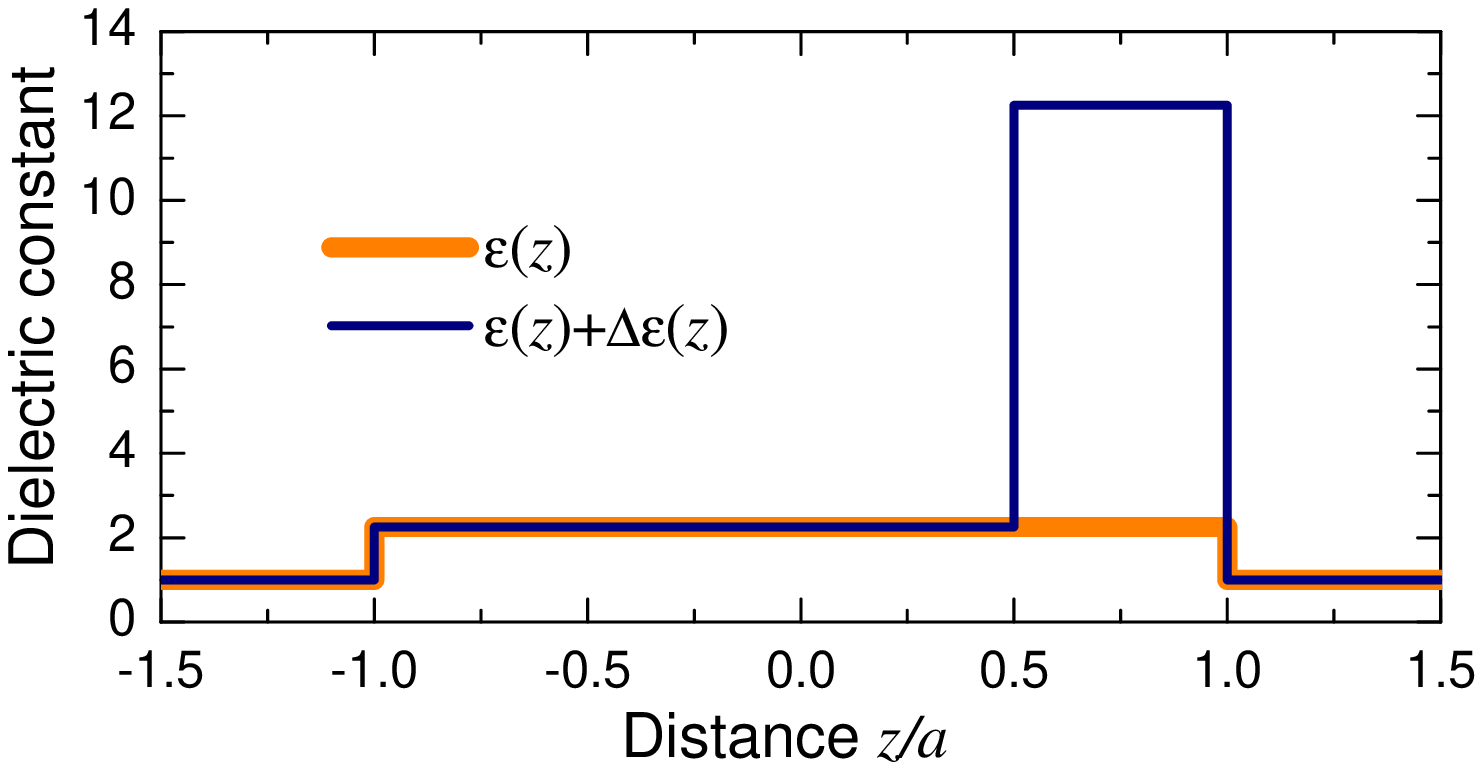}
\caption{Dielectric constants of the unperturbed slab $\varepsilon(z)$ and a slab with a wide
perturbation  $\varepsilon(z)+\Delta\varepsilon(z)$. The distance $z$ is in units of the half width
$a$ of the slab. \label{fig:Diagramp1}}
\end{figure}

The perturbation being considered in this section is given by
\begin{equation}
\Delta\varepsilon(z)=\left\{
\begin{array}{cl}
\Delta\epsilon & \text{for\ \  } a/2 \leq z \leq a\,,\\
0 &\text{otherwise }
\end{array} \right.
\label{pone}
\end{equation}
with $\Delta\epsilon=10$. The profiles of the unperturbed and perturbed dielectric constants are
shown in Fig.\,\ref{fig:Diagramp1}. The analytic solutions of the time-independent Maxwell's
equations using the RS boundary conditions are given in Appendix~\ref{App1}, both for the
unperturbed and the perturbed systems, along with the matrix elements $V_{nm}$ of the perturbation.

\begin{figure}[b]
\includegraphics*[width=0.95\columnwidth]{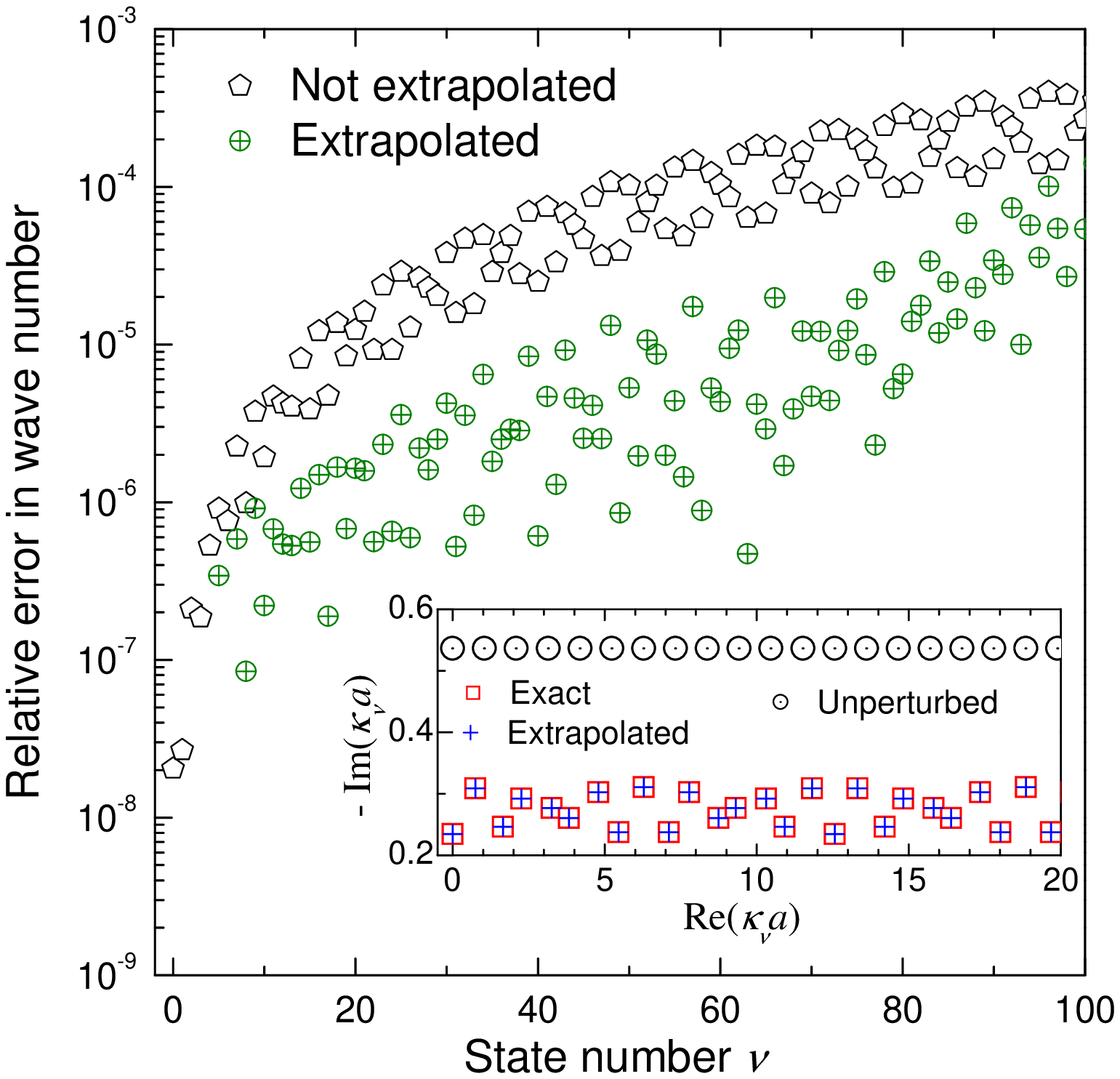}
\caption{Relative errors $\Gamma(\varkappa^{(\infty)}_\nu, \varkappa^{\rm (exact)}_\nu)$ and
$\Gamma(\varkappa^{(N_4)}_\nu, \varkappa^{\rm (exact)}_\nu)$ of the RS wave vectors calculated via
the RSE for the perturbation shown in Fig.\,\ref{fig:Diagramp1}, with and without extrapolation,
respectively, for $N_4=801$. Inset: unperturbed and perturbed RS wave numbers; the latter are
calculated analytically (empty squares) and via the RSE with extrapolation (crosses).
}\label{fig:ErrorP1}
\end{figure}

Using the procedure introduced in Section~\ref{sec-convergence} we calculate four sets of
perturbed wave numbers and extrapolate $\varkappa_{\nu}$ according to Eq.\,(\ref{Improved}). We
also calculate the exact wave numbers ${\varkappa^{\rm{(exact)}}_{\nu}}$ and match up exact and
perturbed states using the MA. The resulting exact and extrapolated eigenvalues
${\varkappa^{{(\infty)}}_{\nu}}$ are shown in the inset of Fig.\,\ref{fig:ErrorP1}, together
with the unperturbed wave vectors. We measure the errors in ${\varkappa^{{(\infty)}}_{\nu}}$
relative to ${\varkappa^{\rm{(exact)}}_{\nu}}$ by $\Gamma(\varkappa^{ (\infty)}_\nu,
\varkappa^{\rm (exact)}_\nu)$ and compare it with  $\Gamma(\varkappa^{(N_4)}_\nu, \varkappa^{\rm
(exact)}_\nu)$ to evaluate the extrapolation method. The results are shown in
Fig.\,\ref{fig:ErrorP1}. We see that the relative error of the RS wave number is generally reduced
by extrapolation by more than one order of magnitude.

\begin{figure}[t]
\includegraphics[width=\columnwidth]{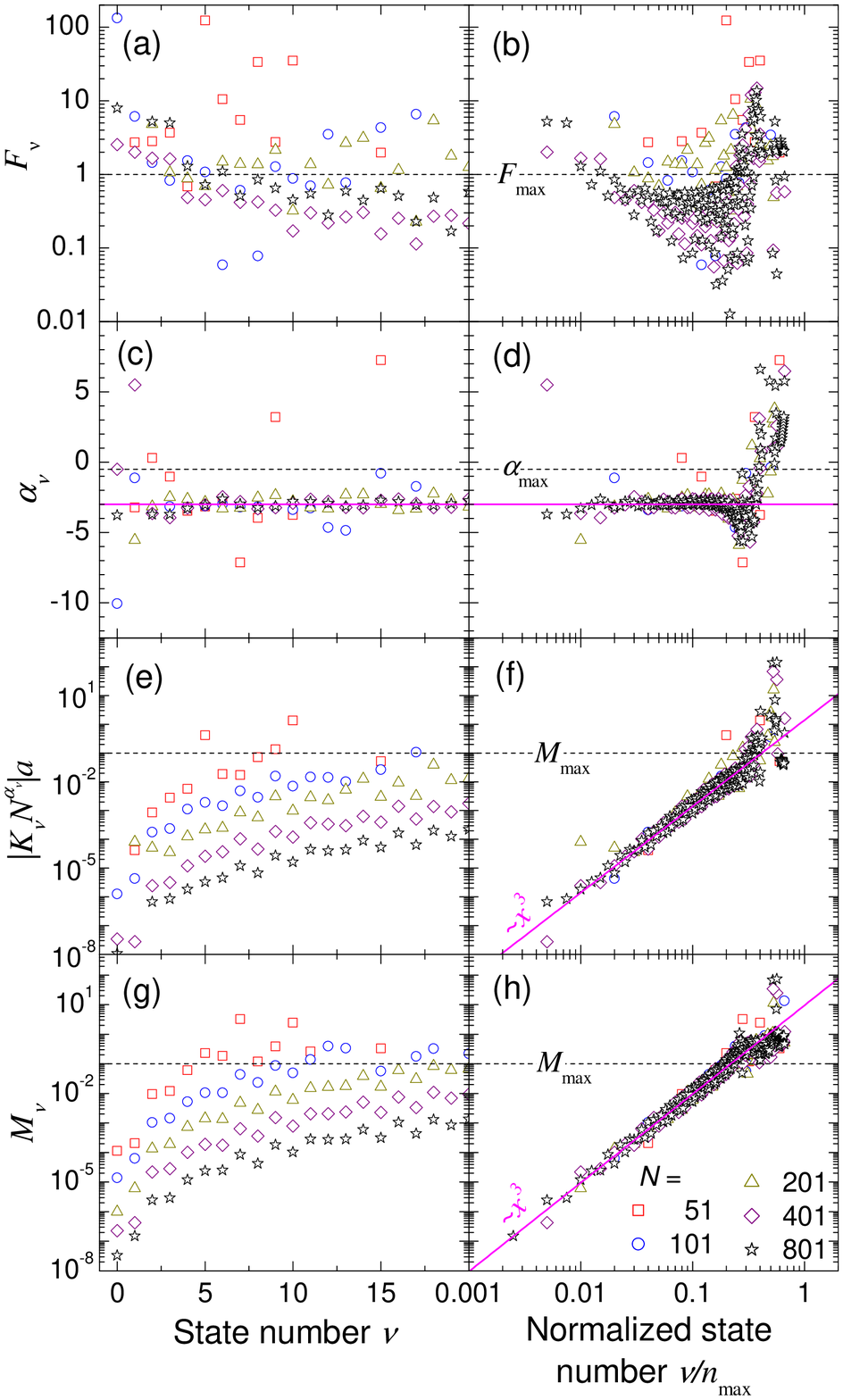}
\caption{Power law parameters and error estimates for the wide perturbation. (a),(b): Relative
extrapolation error $F_\nu$. (c),(d): exponent $\alpha_\nu$ in the power law fit. (e),(f): absolute
errors $K_\nu N^{\alpha_\nu}$ and (g),(h): $M_\nu$ as function of the the state number $\nu$,
calculated for different basis size $N$. The right panels display the data versus the state number
$\nu$ normalized to its maximum value $n_{\rm max}=(N-1)/2$. Straight magenta lines are $\alpha=-3$
(c),(d) and power law fits (h),(f).} \label{fig:PowerlawP1}
\end{figure}

The coefficients and exponents of the power law fit give us information about the convergence
properties of the perturbed RS. For the wide perturbed layer they are shown in
Fig.\,\ref{fig:PowerlawP1}. We see in Fig.\,\ref{fig:PowerlawP1}\,(a) that states close to the
origin in complex wave number space (and having small state number values) are not described well by the power law ($F_{\nu}$ is larger than $F_{\rm{max}}$), even though Fig.\,\ref{fig:ErrorP1} suggests that these states are well
converged. This is reflected in the small absolute error $M_{\nu}$ shown in
Fig.~\ref{fig:PowerlawP1}(g),(h), passing the SC. We also see that for higher wave-number states passing the relative SC
the exponent in the power law is close to $\alpha=-3$ [horizontal lines in Fig.\,\ref{fig:PowerlawP1}\,(c),(d)], in accordance with the
findings in Ref.\,\onlinecite{Muljarov10}. 

Furthermore, the absolute errors $K_\nu N^{\alpha_\nu}$ and $M_\nu$ show universal dependencies on
the normalized state number $\nu/n_{\rm max}$, as shown in
Fig.\,\ref{fig:PowerlawP1}\,(f) and (h). This provides us with a scaling law of the absolute errors
versus the state number:
\begin{equation}\label{Mpower}
M_\nu\propto (\nu/N)^3\,.
\end{equation}
This cubic scaling is shown in Fig.\,\ref{fig:PowerlawP1}\,(f),(h) by straight magenta lines. The
power law exponent $\alpha$ also shows a universal dependency on the normalized state number, being
$\alpha =-3$ for $\nu/n_{\max}\lesssim 0.2$ as can be seen in Fig.\,\ref{fig:PowerlawP1}\,(d). In this region the states pass the relative SC and are extrapolated.

\begin{figure}[t]
\includegraphics*[width=0.95\columnwidth]{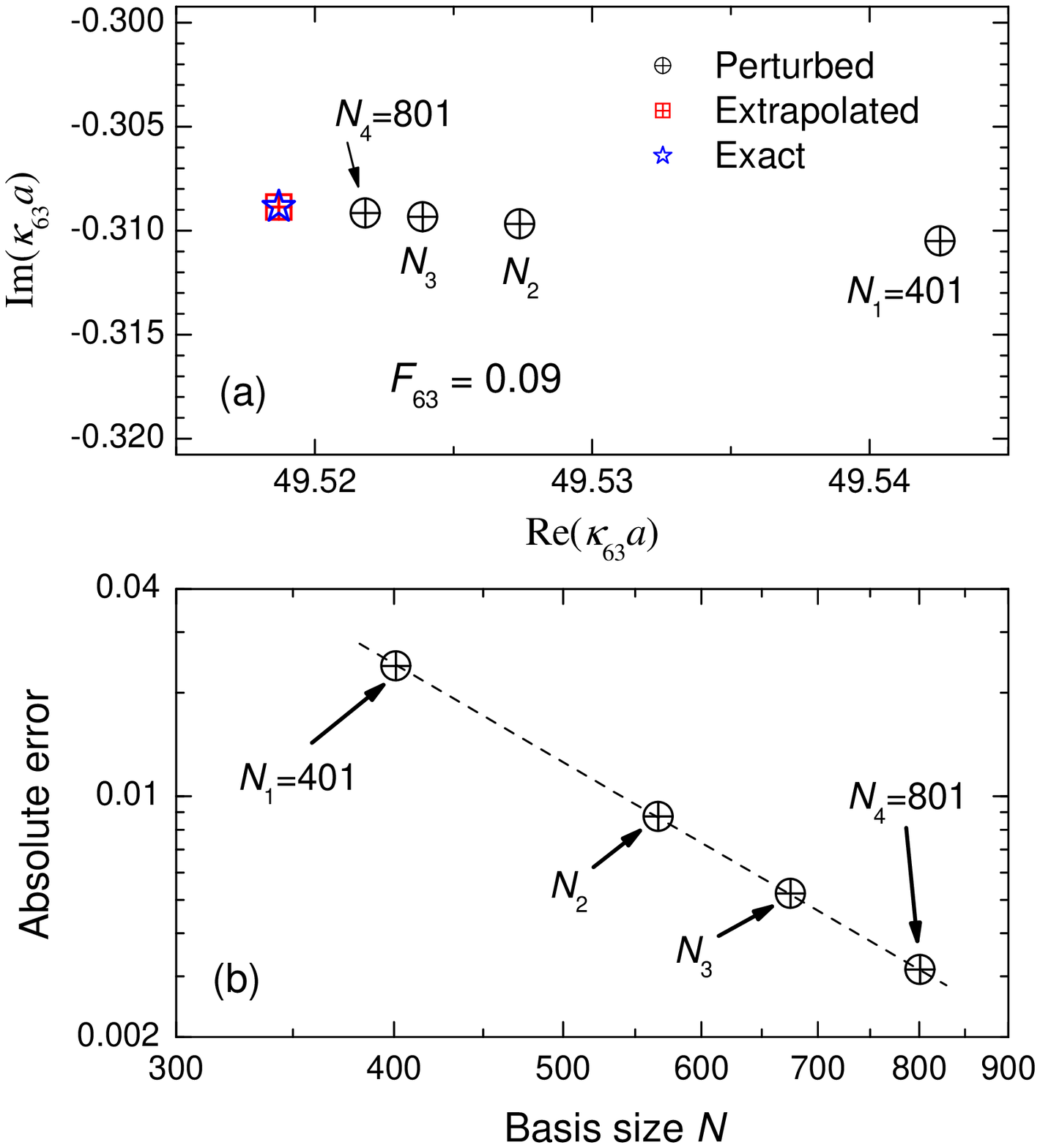}
\caption{(a) Wave number of the perturbed state $\nu=63$ calculated with different basis sizes $N$
and extrapolated to the exact value. (b) Absolute ``exact'' error $|\varkappa^{\rm
(exact)}_{63}-\varkappa^{(N)}_{63}|$  for different $N$  (squares) and a power law fit (dashed
line).} \label{fig:DEMOPL}
\end{figure}

An example of how the power law is applied to extrapolate the wave number of a particular state
$\nu=63$ is given in Fig.\,\ref{fig:DEMOPL}\,(a). Clearly, the
extrapolation leads to a considerable improvement of the accuracy compared to wave number
calculated with the maximum matrix size $N_4$. This is due to the good power law convergence as
shown in Figure~\ref{fig:DEMOPL}\,(b), seen by the straight line connecting the ``exact'' errors
$|\varkappa^{\rm (exact)}_\nu-\varkappa^{(N_i)}_\nu|$ for the four basis sizes.

\begin{figure}[t]
\includegraphics*[width=0.95\columnwidth]{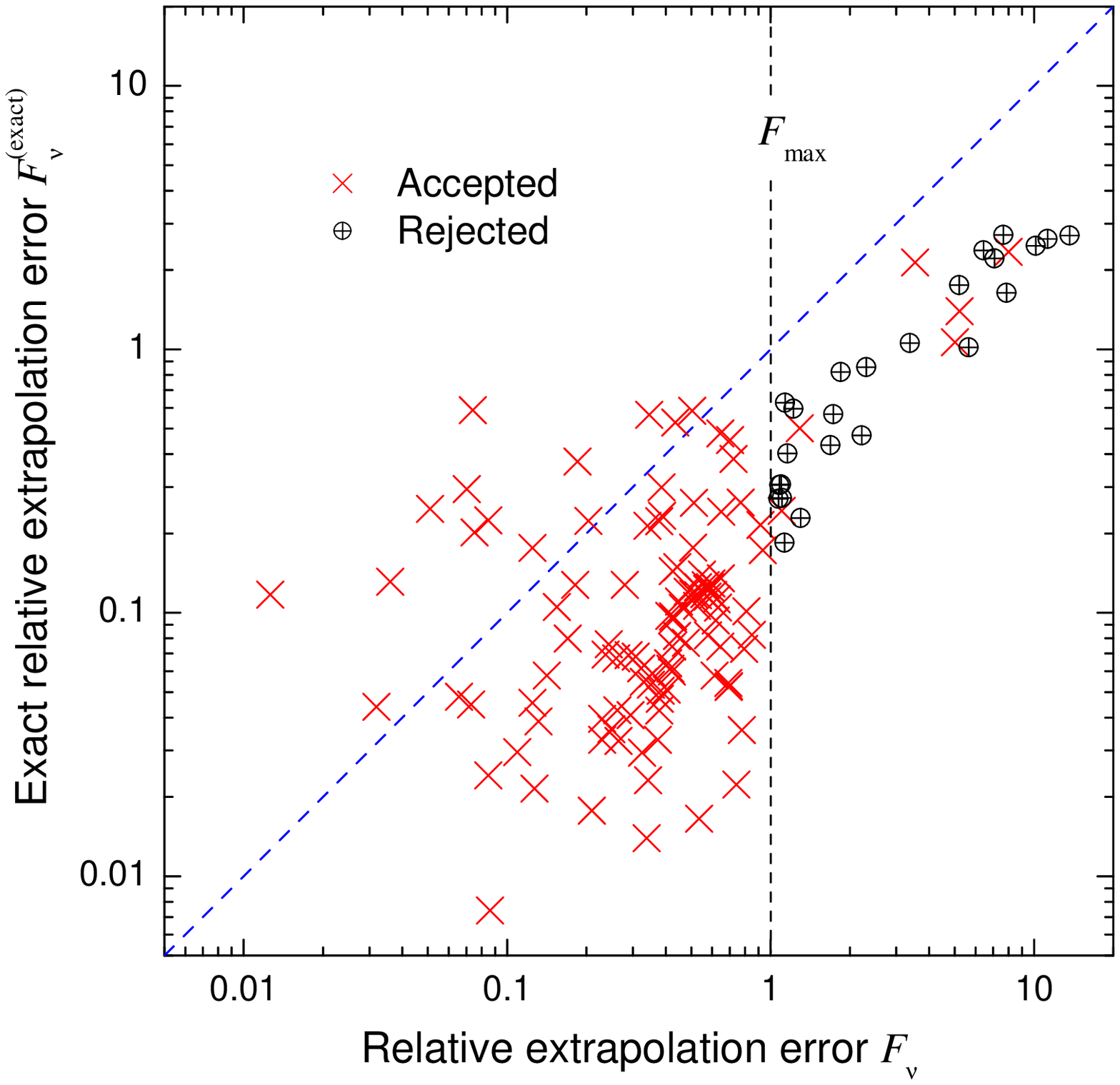}
\caption{ ``Exact'' relative extrapolation error $F_\nu^{\rm (exact)}$ versus relative
extrapolation error $F_\nu$ for both accepted and rejected states, for $N_4=801$. The blue dashed
line shows the anticipated behavior $F_\nu^{\rm (exact)}\approx F_\nu$.} \label{fig:MvsFit}
\end{figure}
 The exact errors are only available if the exact solution is known, but in this ideal case we do not
need the RSE. In a realistic case for which no such solution is known, we need to estimate the
error of the power law extrapolation, which we do using the extrapolation SC and Eq.\,(\ref{Quality}). In order to check how good this estimation is, we compare $F_\nu$ with the
exact relative extrapolation error $F_\nu^{\rm (exact)}=\Phi(K_\nu N^{\alpha_\nu}_4, \varkappa^{\rm
(exact)}_\nu-\varkappa^{(N_4)}_\nu)$. Such a comparison is shown in Fig.\,\ref{fig:MvsFit} for all
states with $\alpha_\nu<-0.5$. We can see that the exact error $F_\nu^{\rm (exact)}$ is typically
overestimated by $F_\nu$, and for all states with $F_\nu<F_{\rm max}$ we have $F_\nu^{\rm
(exact)}<1$, {\em i.e.} the extrapolation is improving the error.  $F_\nu$ can thus be used
reliably to verify the convergency and power law extrapolation.

\subsection{Electric fields}

\begin{figure}[t]
\includegraphics*[width=0.95\columnwidth]{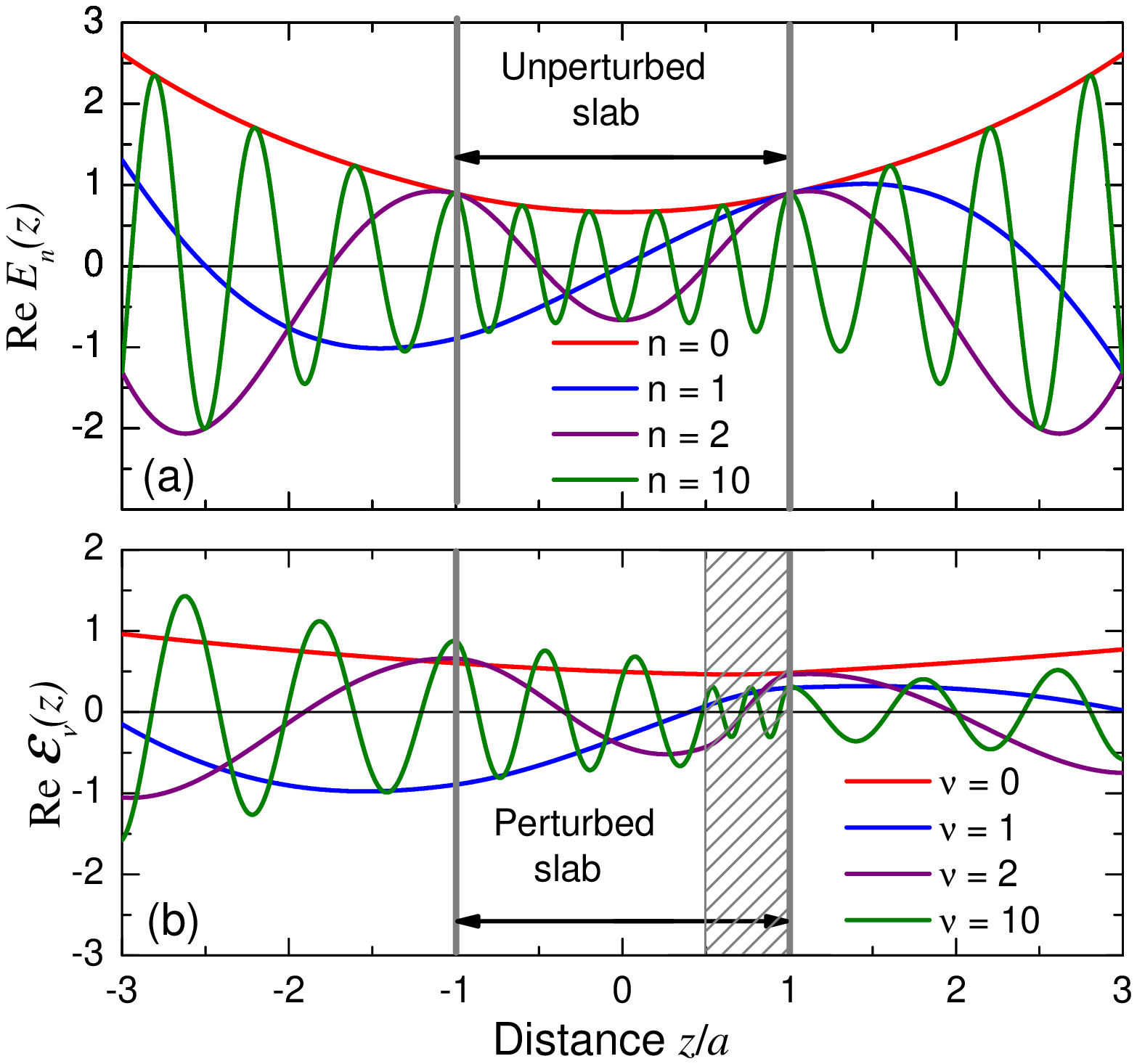}
\caption{Real part of the normalized electric field of a few lowest energy RS of the unperturbed
slab (a) and  of the perturbed slab (b).} \label{fig:statefields}
\end{figure}

The electric fields (EF) $\mathcal{E}_{\nu}(z)$ of the perturbed RS calculated via the exact formula
Eq.\ (\ref{wpl}) are shown in  Fig.\,\ref{fig:statefields} for a few lowest states in comparison with
$E_n(z)$, the EF of the unperturbed RS, given by Eq.\,(\ref{basise1}). The perturbed RS are normalized as in 
Eq.\,(\ref{nint}). In particular, their orthonormality condition reads
\begin{eqnarray}
&&\int_{-a}^{a} \varepsilon_p(z)\mathcal{E}_{\nu}(z)\mathcal{E}_{\mu}(z)\,dz\nonumber
\\
&&-\frac{\mathcal{E}_{\nu} (-a)\mathcal{E}_{\mu} (-a) +\mathcal{E}_{\nu} (a)\mathcal{E}_{\mu}
(a)}{i({\varkappa}_{\nu}+{\varkappa}_{\mu})}=\delta_{\nu\mu}\,,
 \label{nint2}
\end{eqnarray}
where $\varepsilon_p(z)=\varepsilon(z)+\Delta\varepsilon(z)$ is the perturbed dielectric profile.
All unperturbed states have the same imaginary part of their wave vectors (see the inset in
Fig.\,2) and thus their fields have all the same envelope, exponentially growing outside the slab,
with the higher-$n$ states oscillating more rapidly, see Fig.\,\ref{fig:statefields} (a). In the
perturbed system, the envelopes are different due the varying Im\,$\varkappa_\nu$.
Also, as can be seen in Fig.\,\ref{fig:statefields}(b), the frequency of the oscillations increases in the
perturbed (denser) layer, and their amplitudes change at the same time.

\begin{figure}[t]
\includegraphics*[width=\columnwidth]{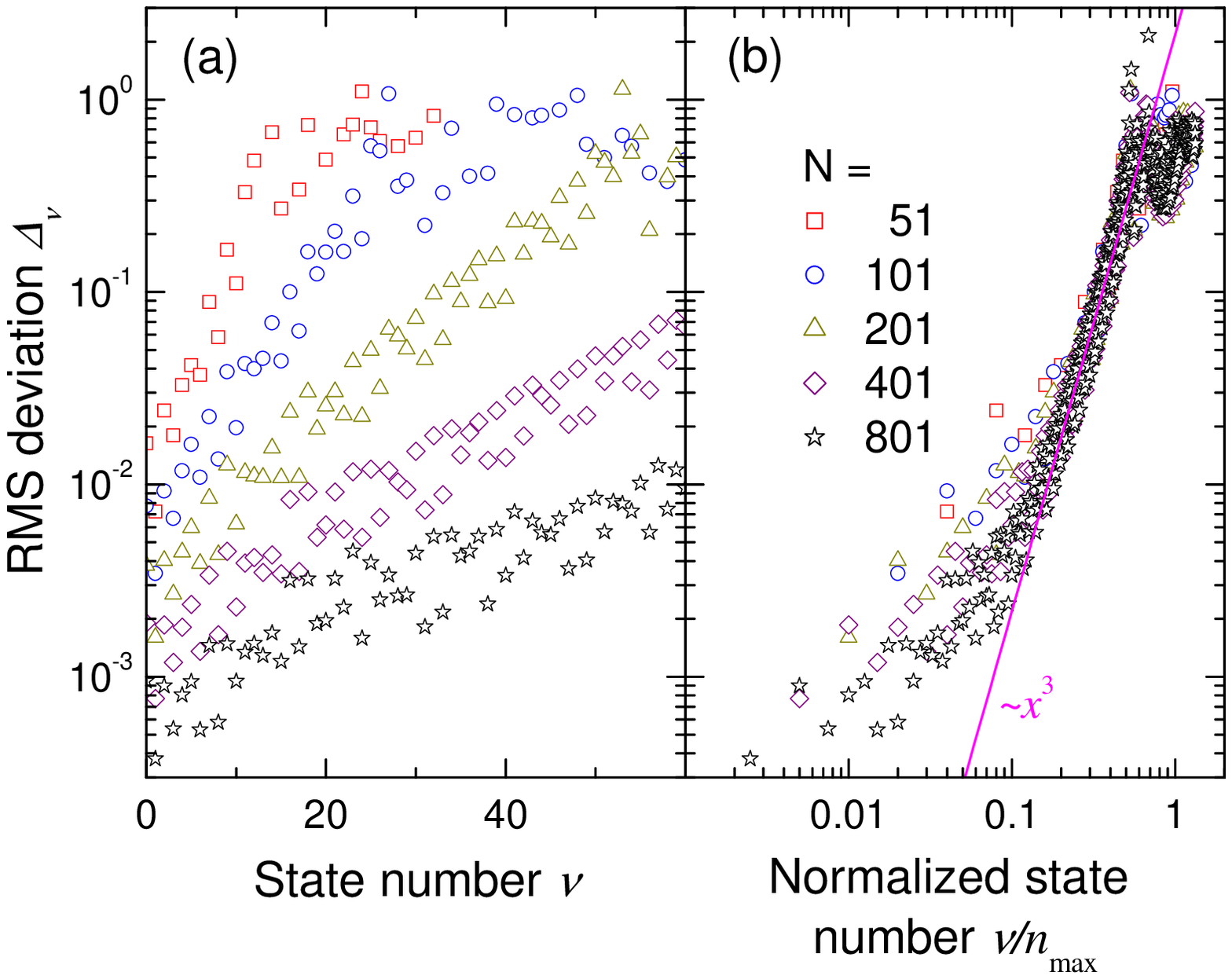}
\caption{Root mean square deviation of the RS electric field $\mathcal{E}^{N}_{\nu}$ from its exact
value $\mathcal{E}^{\rm (exact)} _{\nu}$ versus the state number $\nu$ (a) and  normalized state
number $\nu/n_{\rm{max}}$ (b), calculated for different basis sizes $N$. The straight magenta line
in (b) is a cubic law fit.} \label{fig:figEE}
\end{figure}

The perturbation theory fully reproduces the EF of the RS, both inside and outside the slab. Inside
the slab, the EF is given by the expansion in Eq.\,(\ref{superp}) with the coefficients $c_{n\nu} $
diagonalizing the matrix in Eq.\,(\ref{diag}). Outside the slab, the fields are given by
Eq.\,(\ref{BC}) in which the perturbed wave vectors $\varkappa_\nu$ assign the proper exponential
growth and oscillations of the EF in vacuum, while the amplitudes $A_\nu^\pm$ are found by
comparing Eqs.\,(\ref{BC}) and (\ref{superp}) and using the continuity of the EF through the
boundaries. To quantify how well the perturbation theory reproduces the EF of a RS, we calculate
its root mean square (RMS) deviation within the system defined by
\begin{equation}
\Delta_{\nu}=\sqrt{\frac{\int^{a}_{-a} \bigl|\mathcal{E}^{(N)}_{\nu}(z)-\mathcal{E}^{\rm
(exact)}_{\nu}(z)\bigr|^2\,dz }{\int^{a}_{-a} \bigl|\mathcal{E}^{\rm
(exact)}_{\nu}(z)\bigr|^2\,dz}}\,. \label{rms1}
\end{equation}
The results are shown in Fig.~\ref{fig:figEE}, where we have matched exact and perturbed RS using
the MA and plotted $\Delta_\nu$ for different basis sizes $N$. We see that the trend in accuracy
with state number and the basis size is the same as in Fig.\,\ref{fig:PowerlawP1}(e),(g), and the
RMS deviation versus the normalized state number also shows a universal dependence similar to those
in Fig.\,\ref{fig:PowerlawP1}(f),(h). However, the EF is in general less well reproduced than the
wave numbers and the power law $\Delta_\nu\propto (\nu/N)^3$ is observed only in the interval of
$0.05<\Delta_\nu <0.2$.

\subsection{Green's function and transmission}

The Green's function (GF) is an important quantity which fully characterizes the response of an
optical system, determining its scattering and transmission. For the slab with a wide perturbed
layer given by Eqs.\,(\ref{unpslab}) and (\ref{pone}), the GF $G(z,z';k)$ which satisfies the
equation
\begin{equation}
\Big\{ \partial^{2}_{z}+\bigl[\varepsilon(z)+\Delta\varepsilon(z)\bigr] k^{2}\Bigr\}
G(z,z';k)=\delta(z-z') \label{AGreens}
\end{equation}
and outgoing boundary conditions can be calculated analytically. Note that when calculating
observables, $k$ is real as it is given by the vacuum wave number of an external driving field. The 
GF is calculated using its spectral representation,\cite{Newton60,More71,Muljarov10}
\begin{equation}
G(z,z';k)=\sum_{\nu}\frac{\mathcal{E}_{\nu}(z)\mathcal{E}_{\nu}(z')}{2k(k-{\varkappa}_{\nu})}\,,
\label{MLtheory}
\end{equation}
in which the EF $\mathcal{E}_{\nu}(z)$ and the RS wave numbers $\varkappa_{\nu}$ are calculated
numerically via the RSE. For the wave numbers $\varkappa_{\nu}$, we use the extrapolated values
Eq.\ (\ref{Improved}). 

\begin{figure}[t]
\includegraphics*[width=0.95\columnwidth]{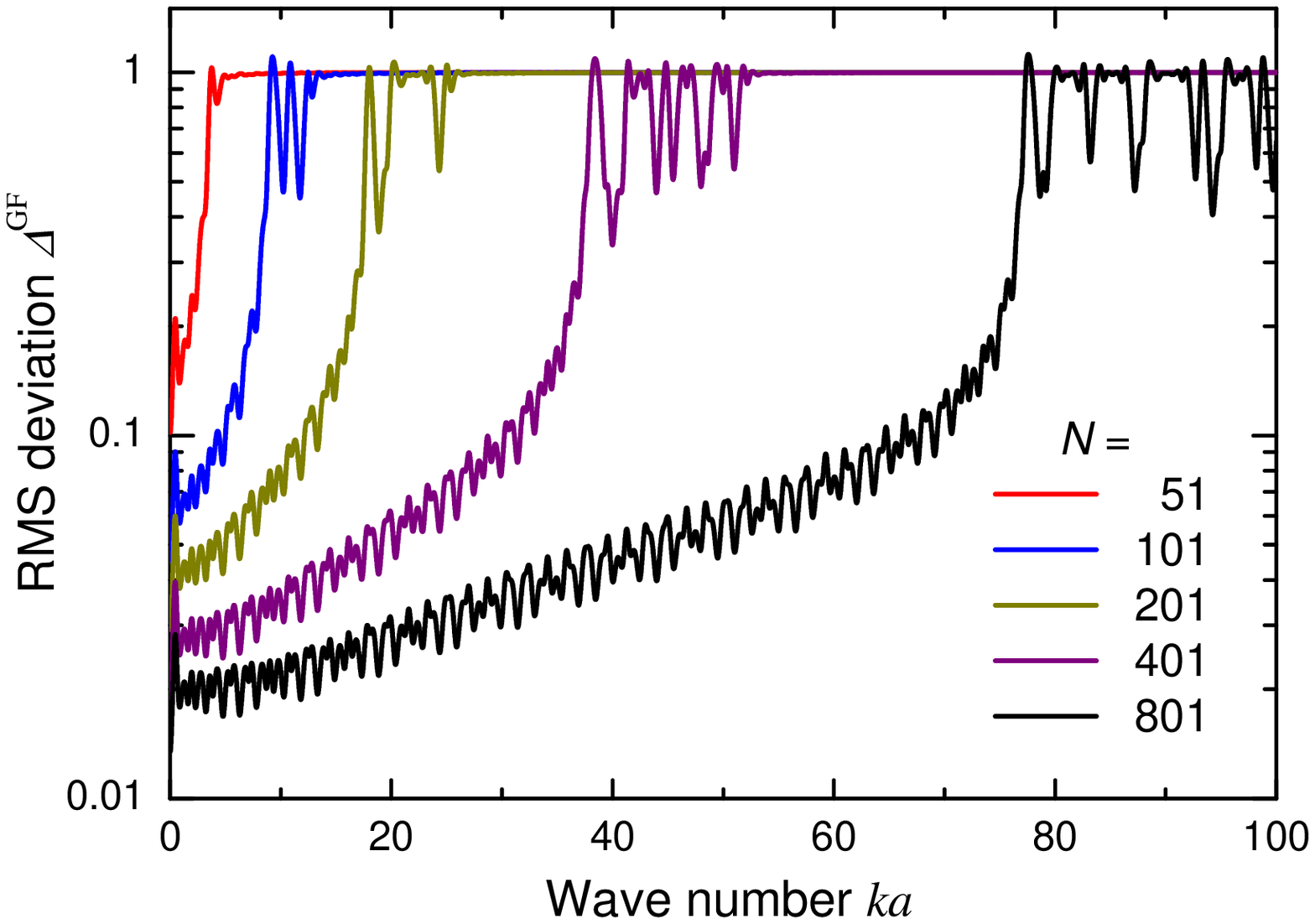}
\caption{The root mean square deviation in the GF $\Delta^{\rm{GF}}$ as a function of the wave
number of the driving field $k$, calculated via the RSE for different basis sizes $N$.}
\label{fig:GreensA}
\end{figure}

In light of the importance of the GF and its further usage for calculation of observables, we
compare $G^{(N)}$, the GF calculated by RSE with basis size $N$ and Eq.\,(\ref{MLtheory}), to its
exact analytic form $G^{\rm{(exact)}}$, again using the RMS deviation as given by
\begin{equation}
\Delta^{\rm{GF}}=\sqrt{\frac{\int^{a}_{-a}\!\int^{a}_{-a} \bigl|
G^{(N)}(z,z')-G^{\rm{(exact)}}(z,z')\bigr|^{2}\,dz dz'}{\int^{a}_{-a}\!\int^{a}_{-a} \bigl|
G^{\rm{(exact)}}(z,z')\bigr|^{2}\,dz dz'}}\,. \label{RMS2}
\end{equation}
Such a comparison is shown in Fig.\,\ref{fig:GreensA} for different basis sizes $N$. Increasing the
basis size has two effects on the GF: (i) it improves the GF error at a given $k$ and (ii) widens
the $k$-range of the GF with small error. The latter is due to a larger wave-number range of poles
in the GF, Eq.\,(\ref{MLtheory}), being reproduced for large $N$.

Both expansions Eqs.\,(\ref{superp}) and (\ref{MLtheory}), for the EF and for the GF, are valid
only inside the slab or on its borders and are not suitable for the vacuum area where the EF of the
RS grow exponentially. The GF itself is, however, regular on the real $k$-axis. Moreover, in
vacuum, it always has a simple analytic form of a plane wave with the amplitude that can be deduced
from values inside the slab,  Eq.\,(\ref{MLtheory}), using the continuity of the GF when passing
through the interfaces. In this way, the GF can be calculated at any point of the $(z,z')$ space,
inside or outside the slab.

\begin{figure}[t]
\includegraphics*[width=0.95\columnwidth]{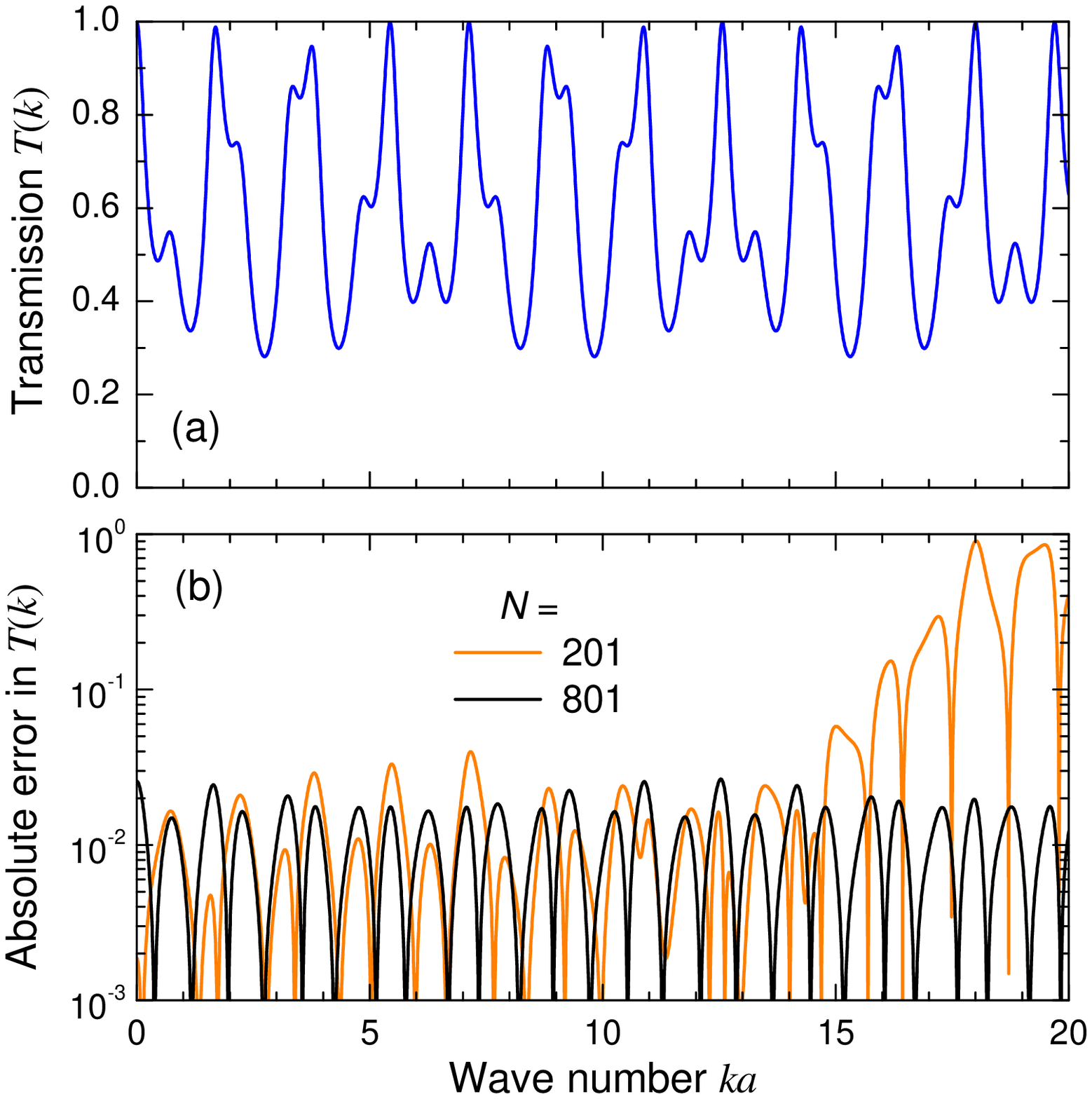}
\caption{(a) Light transmission through the slab with a wide-layer perturbation Eq.\,(\ref{pone}).
(b) Absolute error in the transmission calculated using the analytic form of $T(k)$ and numerical
values from the RSE for two different simulations.}
\label{fig:abserrO}
\end{figure}

The delta-function in Eq.\,(\ref{AGreens}) plays the role of a source of plane waves generated at
the point $z'$ and propagating in both directions, away from the source. The GF then has the
meaning of the system's response on such a plane-wave excitation. This can be used to derive a
formula for the transmission in terms of the GF. To do this, we place the source of strength $2ik$
just outside the slab at $z'=-a$, in order to produces two plane wave of amplitude 1. One of these
waves is transmitted trough the slab, and just after the slab at point $z=a$ the intensity of the
EF (which does not change with further increase of $z$) is given by
\begin{equation}
T(k)=\left|2kG(a,-a;k)\right|^{2} \label{Trans1}
\end{equation}
and is called transmission.

We calculate the transmission using Eqs.\,(\ref{MLtheory}) and (\ref{Trans1}) for the GF taken to
be either numerical $G^{(N)}$ or analytical $G^{\rm{(exact)}}$. This allows us to calculate the
absolute error in the transmission, $|T^{(N)}-T^{\rm (exact)}|$, which is shown in
Fig.\,\ref{fig:abserrO}(b). The transmission itself is shown in Fig.\,\ref{fig:abserrO}(a) and has
a profile which is fully determined by the pole structure of the GF. The RS which contribute in
this frequency range can be seen in the inset to Fig.\,2. We see that the error of the transmission
has a similar magnitude and scaling with $N$ as the GF itself, as can be expected from
Eq.\,(\ref{Trans1}).

\subsection{$\delta$-perturbation}

\begin{figure}[t]
\includegraphics[width=0.9\columnwidth]{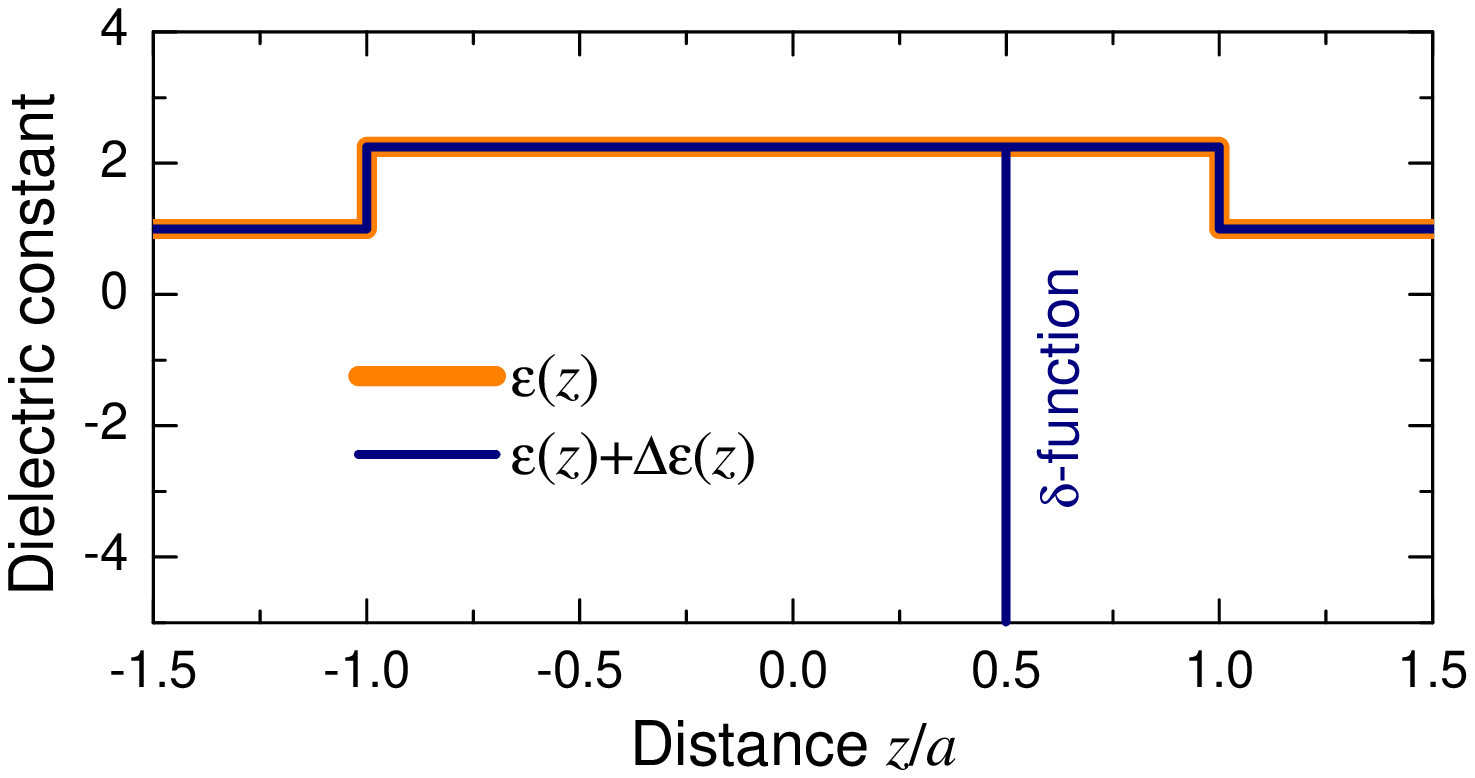}
\caption{Dielectric constants of the unperturbed slab $\varepsilon(z)$ and a slab with a
$\delta$-perturbation  $\varepsilon(z)+\Delta\varepsilon(z)$. The distance $z$ is in units
of the half width $a$ of the slab.}
\label{fig:Diagramp2}
\end{figure}

We now move from a wide perturbation to a very narrow and strong one, like a thin metal film on a dielectric. Such a perturbation is described by
\begin{equation}
\Delta\varepsilon(z)=w\epsilon_d\delta(z-a/2)
    \label{ptwo}
\end{equation}
with the delta-scatter strength $w\epsilon_d=-0.1a$. Physically, this perturbation corresponds to a
thin layer of the dielectric constant changed by $\epsilon_d$, which is placed at $z=a/2$ and has
a width $w$ much narrower than the shortest wavelength of the resonant modes used in the basis. The
dielectric profile for the system  with the $\delta$-perturbation is shown in
Fig.~\ref{fig:Diagramp2}.

\begin{figure}[b]
\includegraphics[width=0.95\columnwidth]{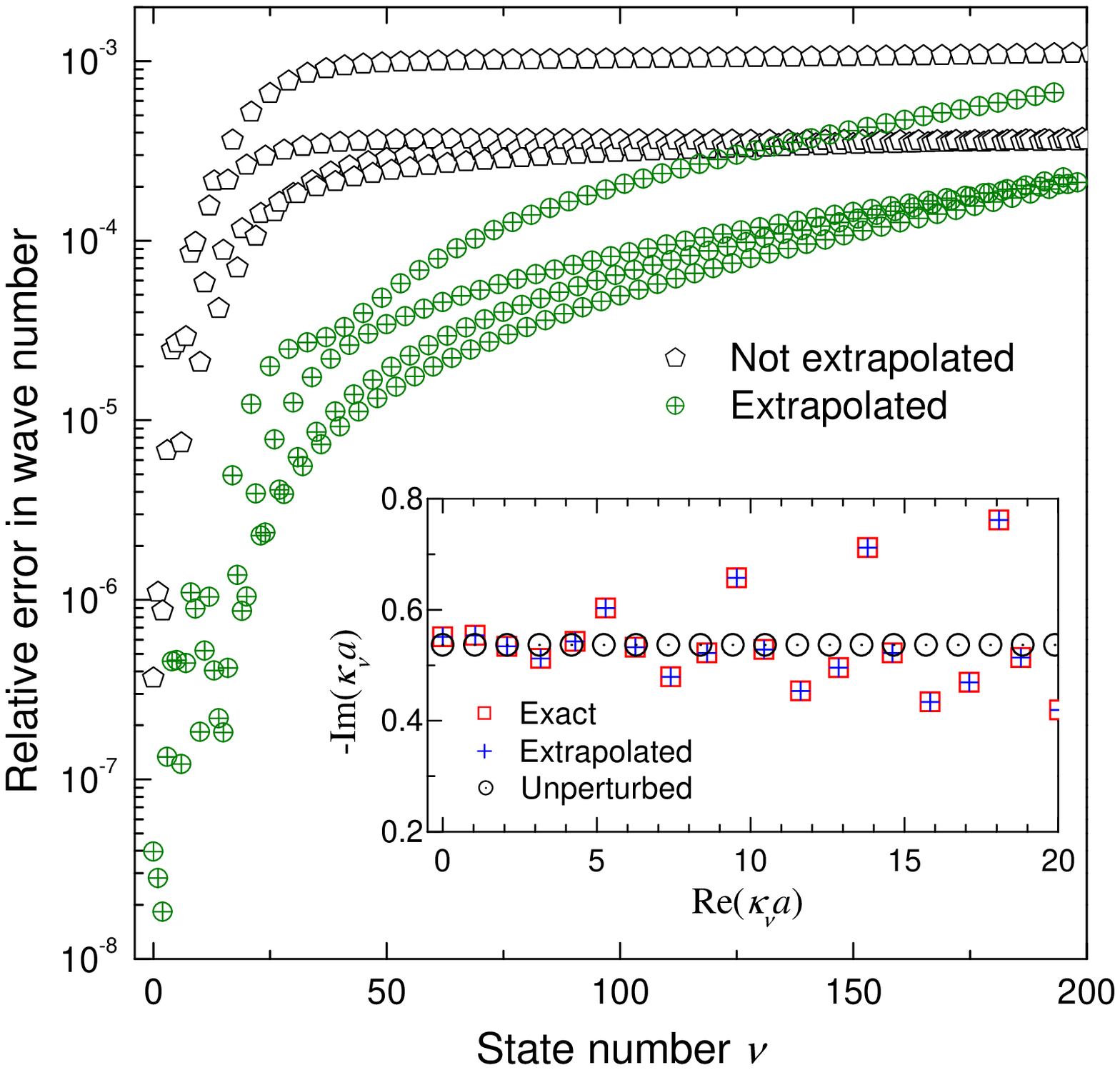}
\caption{As Fig.\,2, but for the $\delta$-perturbation shown in Fig.\,\ref{fig:Diagramp2}.}
\label{fig:ErrorP2}
\end{figure}

As in the case of a wide-layer perturbation considered in Section~\ref{widelayer} we plot and compare
in Figs.\,11--14 the RS wave numbers, calculated exactly and via the RSE with and without
extrapolation, as well as the parameters of the power law fit and relative and absolute errors
which we also need for the quality check of our simulation and extrapolation. The analytic
solutions for the $\delta$-perturbation and its matrix elements are given in the
Appendix \ref{App1}.

\begin{figure}[b]
\includegraphics[width=\columnwidth]{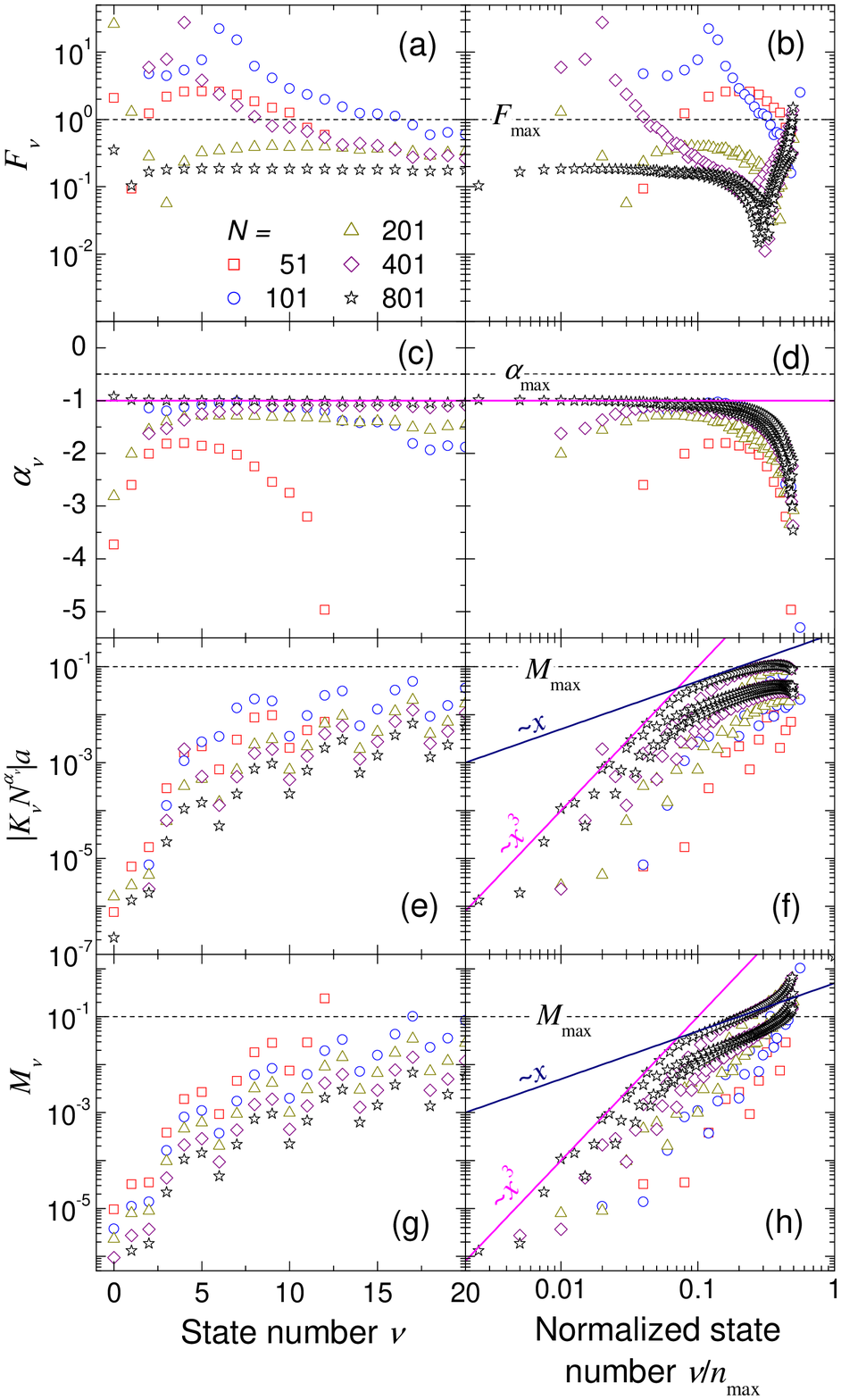}
\caption{As Fig.\,3 but for the $\delta$-perturbation shown in Fig.\,\ref{fig:Diagramp2}.
The horizontal magenta lines in panels (c) and (d) are $\alpha=-1$ lines. }
\label{fig:PowerlawP2}
\end{figure}

We see in Fig.~\ref{fig:ErrorP2} that the extrapolation reduces the relative error by 1-2 orders
of magnitude. The integral strength of the perturbation is much (almost two orders of magnitude) weaker than
in the case of the wide layer considered in Section~\ref{widelayer}. However, the convergence is much slower in the case
of the $\delta$-perturbation. We see in Fig.\,\ref{fig:PowerlawP2}(c),(d) that for large $N$
the power law exponent is close to $\alpha_\nu=-1$. 
This is to be expected as the $\delta$-perturbation does not have a finite width. The matrix elements $V_{nm}$, though oscillating, have no decrease with increasing wave number (or index $n$) which leads to a much stronger mixing of states compared to the wide layer perturbation. Indeed, in the wide layer case, states with higher indices are less important due to the rapid oscillation of their wave functions, so that the matrix elements scale as $V_{nm}\propto 1/n$ (for $n\gg m$). Using the second-order Rayleigh-Schr\"odinger perturbation theory and the explicit form Eqs.\,(\ref{Vnm1}) and (\ref{Vnm2}) of the matrix elements $V_{nm}$, we can show that the wave number corrections scale as $1/N$ and $1/N^3$ for the $\delta$- and wide-layer perturbations, respectively, in accordance with Figs.\,3(d) and 12(d).

\begin{figure}[t]
\includegraphics*[width=0.9\columnwidth]{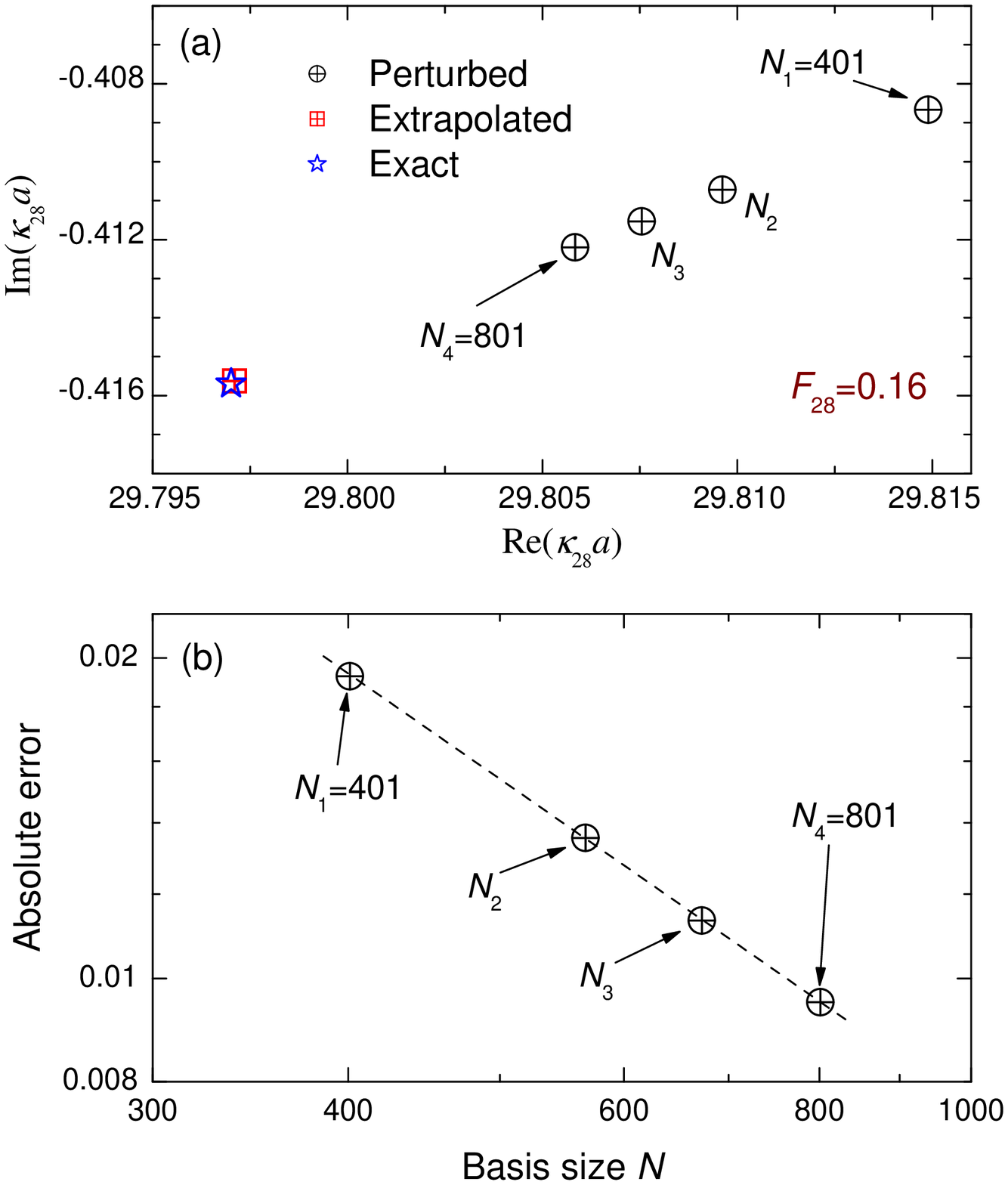}
\caption{As Fig.\,4, but for the $\delta$-perturbation shown in Fig.\,10 and state number $\nu=28$.}
\label{fig:DEMOPL2}
\end{figure}

In the case of the $\delta$-perturbation, the absolute errors shown in Fig.\,\ref{fig:PowerlawP2}(f,h) as
functions of the normalized state number do not display any universal curves, still for small $\nu/N$
approaching asymptotically a cubic law in the state number $\nu$ (magenta lines). Thus we conclude
that in this case $M_\nu\propto \nu^3/N$ [compare with Eq.\,(\ref{Mpower})]. At larger values of $\nu/N$ 
this dependence transforms into a linear one, $M_\nu\propto \nu/N$ (blue lines). Because of the slow
($1/N$) convergence, the extrapolation gives a huge improvement as is clear from Fig.\,\ref{fig:DEMOPL2} and
demonstrates its necessity in the particular case of the $\delta$-perturbation.

\begin{figure}[t]
\includegraphics*[width=0.9\columnwidth]{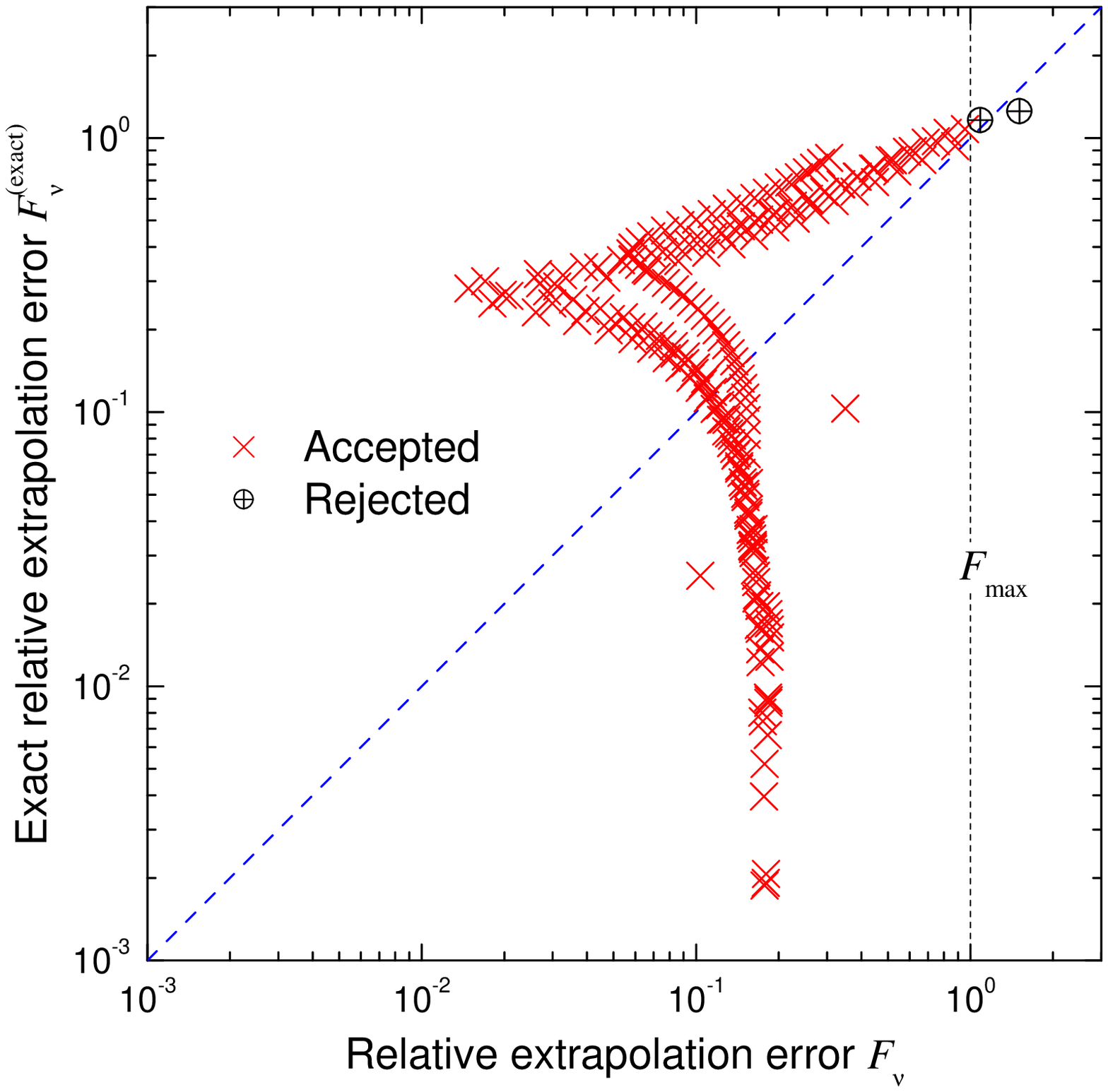}
\caption{As Fig.\,5, but for the $\delta$-perturbation shown in Fig.\,10.}
\label{fig:MvsFit2}
\end{figure}

At the same time, the relative extrapolation error is predicted within an order of magnitude, as
can bee seen in Fig.\,\ref{fig:MvsFit2}. For the majority of RS, $F^{\rm (exact)}_{\nu}<F_{\nu}$,
the exact values $F^{\rm (exact)}_{\nu}$ being significantly overestimated. However, for a large
class of solutions it turns out to be highly underestimated. The systematic deviation seen Fig.\,\ref{fig:MvsFit2} in
estimating the relative extrapolation error though $F_{\nu}$ may be a result of the systematic
variation in the power law exponent $\alpha_{\nu}$ well seen in Fig.\,\ref{fig:PowerlawP2}(c),(d).
Hence it is generally advisable when studying convergence with our method to run simulations with a
variety of $N_4$ parameters in order to establish over what range of $N_4$ the power law is
applicable for the given strength of perturbation. 

We were also able to simulate a $\delta$-perturbation outside the perturbed slab by taking the
unperturbed slab to include the position of the delta scatterer and thus the perturbation
consisting of a superposition of a $\delta$-perturbation and a wide layer compensating the difference in
the dielectric constants between the vacuum and the unperturbed slab. In this case we did obtain
convergence of the perturbed wave numbers to the exact solution. However, for a $\delta$-perturbation
outside of the unperturbed slab or exactly on the border, the simulation does not
converge to the correct solution. This is to be expected since in this case the perturbed RS contain waves reflected from the external perturbation,
which are waves propagating towards the slab. Such incoming waves are not part of the basis of unperturbed RS,
and thus cannot be reproduced by an expansion in this basis.

\subsection{Microcavity}\label{subsec:microcavity}
\begin{figure}[t]
\includegraphics[width=0.9\columnwidth]{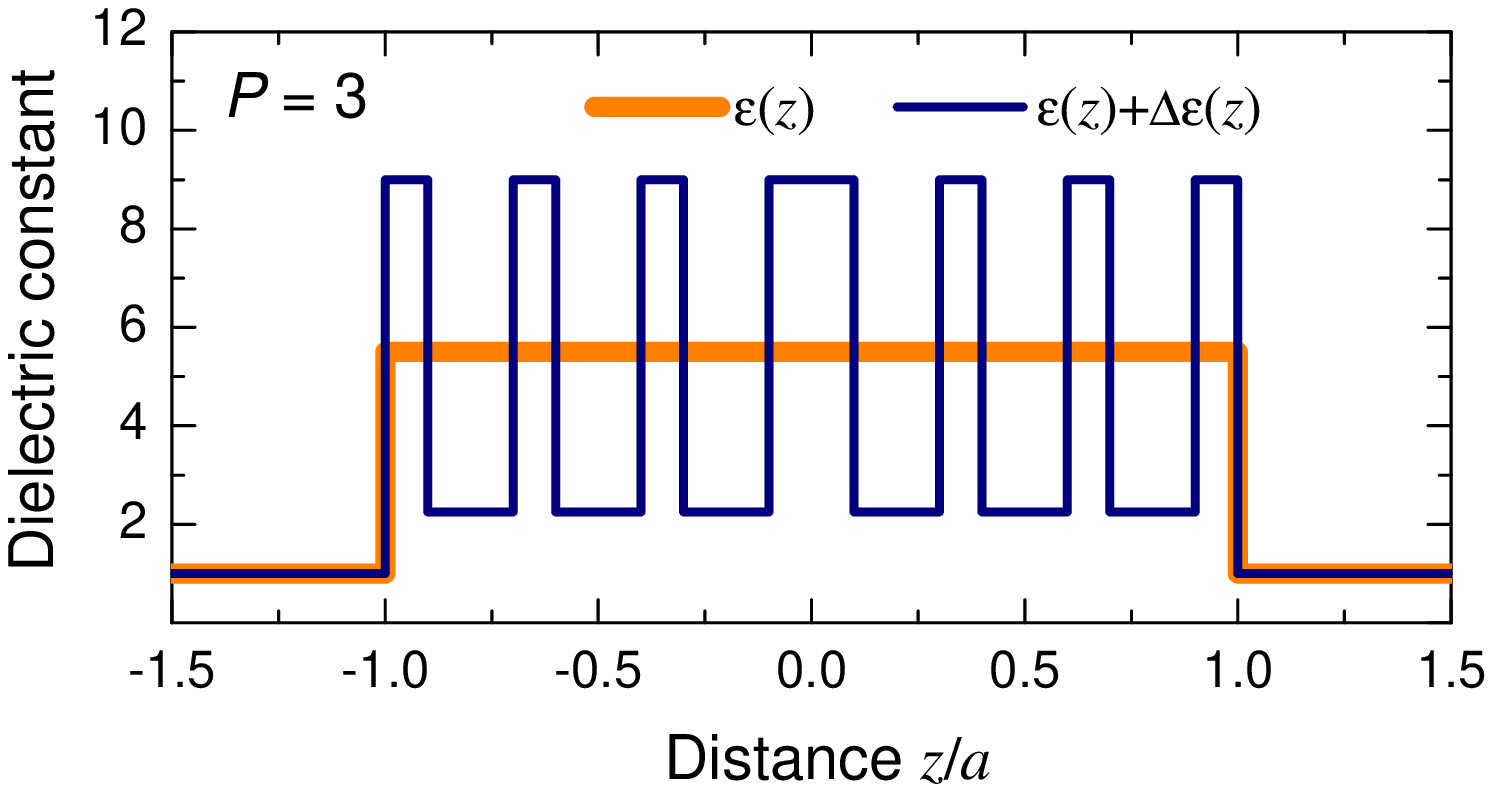}
\caption{Dielectric profiles of a planar microcavity having $P=3$ pairs of Bragg mirrors on each side (blue line) and an unperturbed dielectric slab (orange line).}
\label{fig:Diagramp4}
\end{figure}

To evaluate the RSE in presence of sharp resonances, we use a Bragg-mirror microcavity (MC), which consists of a Fabry-P\'erot cavity of thickness $L_{C}$ and
refractive index $n_{C}$ surrounded by distributed Bragg reflectors (DBRs). The DBRs consist of $P$ pairs of dielectric layers with
alternating high ($n_{H}=3.0$) and low ($n_{L}=1.5$) refractive index. In order to a have sharp cavity mode at a given wavelength $\lambda_{C}$, these alternating
layers have to be of quarter wavelength optical thickness, and the optical thickness of the cavity has to be a multiple of half the wavelength.
We take $L_{C}=\lambda_{C}/2$. An example of the dielectric profile of such
a system with $P=3$ is shown in Fig.\,\ref{fig:Diagramp4}.
\begin{figure}[t]
\includegraphics*[width=0.95\columnwidth]{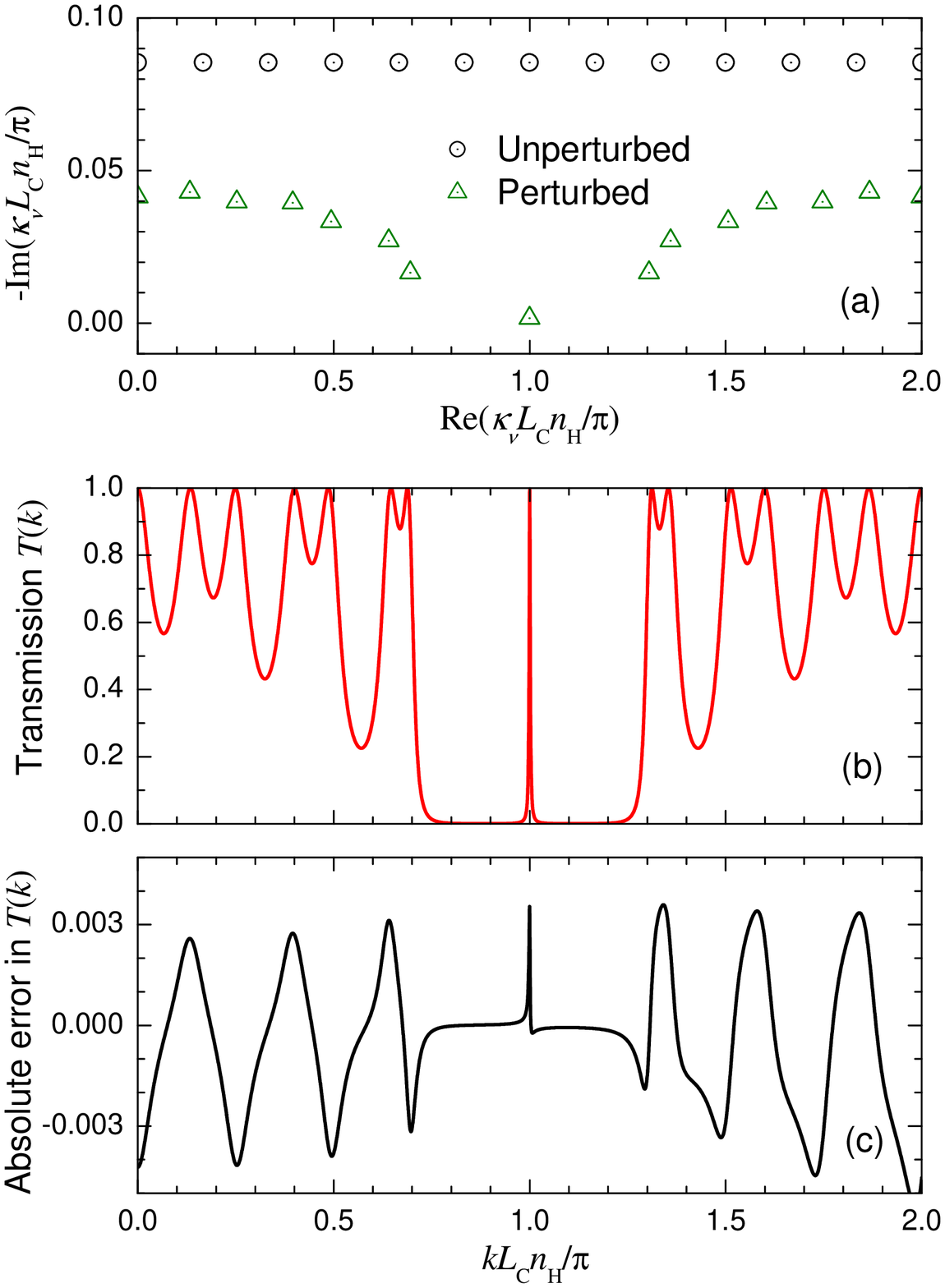}
\caption{(a) Wave vectors $\varkappa_\nu$ of the resonant states of a microcavity with $P=3$ pairs of
Bragg mirrors on each side calculated via RSE with $N=801$.  (b) Microcavity transmission as a function
of the normalized wave vector of the incoming light; $L_{C}$ and $n_{C}$ are the cavity thickness and refractive index.
(c) The difference in the transmission calculated via RSE and using the scattering matrix method.\cite{Tikhodeev02}}
\label{fig:Demo}
\end{figure}
\begin{figure}[t]
\includegraphics*[width=\columnwidth]{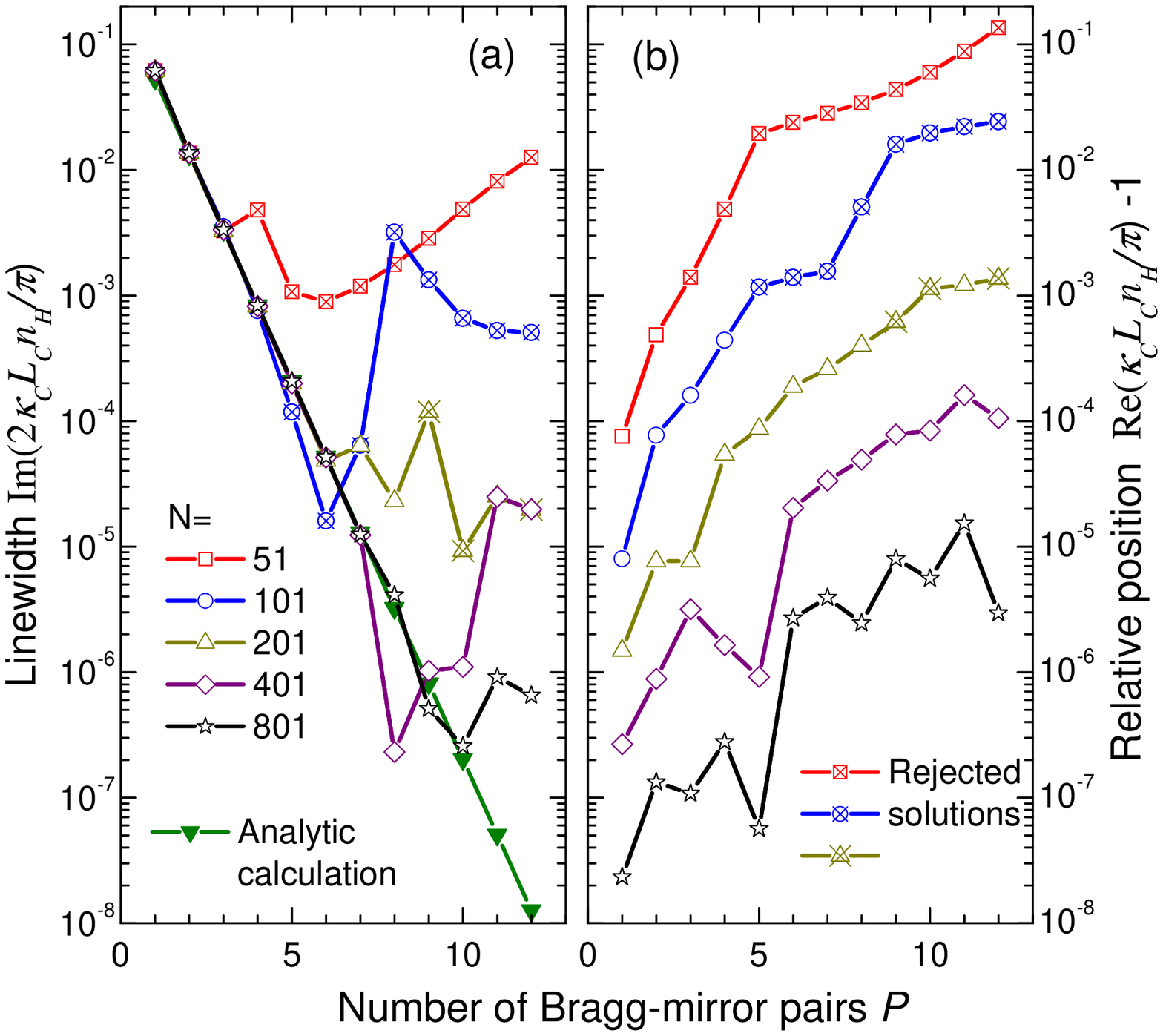}
\caption{The FWHM (a) and the position of the cavity mode (b) calculated analytically and via the
RSE for different number of pairs $P$ of Bragg mirrors on each side of the microcavity.
$N$ is the basis size used in the RSE. Where possible, extrapolated wave numbers have been used.
Crossed rectangles for $N=51$ indicate states which are rejected by the SC.}
\label{fig:FWHM}
\end{figure}

The RS of a MC are calculated using the RSE. The RS wave vectors and the transmission through the
MC are shown in Fig.\,\ref{fig:Demo}(a),(b). For reference, the unperturbed eigenvalues are also
included in Fig.\,\ref{fig:Demo}(a). The unperturbed system taken for the RSE is again a dielectric
slab which dielectric constant $\varepsilon(z)$ can be seen in Fig.\,\ref{fig:Diagramp4}.
Throughout this section the outer boundaries of the MC and the unperturbed slab coincide, and we
choose $\epsilon_s=5.5$ which is between $n_L^2$ and $n_H^2$, providing good convergence. $\epsilon_s$ could be further
optimized for best convergence of the RSE. In order to
verify the transmission calculated by the RSE, we use the scattering matrix
method\cite{Tikhodeev02} which is a straightforward and precise way of calculating the
optical properties of a planar system. Figure~\ref{fig:Demo}(c) demonstrates a good agreement
between the two calculations.

Clearly, there is a one-to-one correspondence between the RS wave vectors in
Fig.\,\ref{fig:Demo}(a)  and the MC transmission in Fig.\,\ref{fig:Demo}(b). Namely, the real part
of the wave vectors corresponds to the positions of the peaks in the transmission while the
imaginary part gives their line widths. This is well understood in view of the spectral
representation of the Green's function Eq.\,(\ref{MLtheory}) used for the calculation of the
transmission via Eq.\,(\ref{Trans1}).

One of the modes shown in Fig.\,\ref{fig:Demo}(a) is rather isolated and has imaginary part much smaller 
than the others. This mode, $\varkappa_{C}$, satisfies the
Fabry-P\'erot resonance condition Re\,$\varkappa_{C}=\pi/(L_{ C} n_{ C})$ and is called the cavity mode.
For the wave vector $k$ of incoming light close to this resonance condition, $k\approx \pi/(L_{ C}
n_{ C})$, the Greens' function  Eq.\,(\ref{MLtheory}) is dominated by a single term corresponding to
this narrow mode. As a consequence, there is a sharp peak in the center of a wide stop-band seen in
the transmission in Fig.\,\ref{fig:Demo}(b). For sufficiently large $P$ an analytic
approximation for its full width at half maximum (FWHM) is known,\cite{Savona95}
\begin{equation}
\Delta k=\frac{4 n_{\rm ext}}{n_{ C}^2}\left(\frac{n_L}{n_H}\right)^{2P}\frac{1}{\displaystyle
L_{ C}+\frac{\lambda_{ C}}{2}\,\frac{n_{ L} n_{ H}}{n_{ C}(n_{ H}-n_{ L})}}\,, \label{AnalyticBM}
\end{equation}
which we use to compare with the RSE calculation. With the refractive index of
the external material $n_{\rm ext}=1$ and using $\lambda_{ C}=2 L_{ C}$ and $n_{ C}=n_{ H}$,
Eq.\,(\ref{AnalyticBM}) reduces to $\Delta k=4(n_{ H}-n_{ L})(n_{ L}/n_{ H})^{2P}/(L_{ C}n_{ H}^3)$. Comparison of
the above formula with the RSE result for the cavity mode is given in
Fig.\,\ref{fig:FWHM}, for different number of Bragg-mirror pairs $P$ and for different basis size
$N$ in the RSE. Figure~\ref{fig:FWHM} demonstrates that RSE is capable of giving both
the correct width and location of sharp resonances in the transmission profile, if a large enough
basis is used, in spite of there being no sharp resonances in the basis. As the basis size is
enlarged, the width and the peak location of the cavity mode converge to the analytic values.
The fact that for a fixed $N$ the cavity mode position and the width are predicted
worse for larger $P$ is explained by our choice of the unperturbed slab which always has exactly
the same thickness as the Bragg-mirror MC. With the number of
Bragg-mirrors increasing, the field inside the MC oscillates more rapidly (also shifting the cavity
mode towards higher frequencies) that requires a larger number of RS to be taken into account in
order to produce results on the same level of accuracy. We have verified (not shown) that 
the errors become independent of $P$, if one and the same constant width of the unperturbed slab is used 
for different values of $P$.

\section{Summary}\label{sec:summary}
The resonant state expansion has been implemented and validated in planar open optical systems reducible
to effective one-dimensional systems. A reliable method of calculation of resonant states, and in
particular their wave numbers, electric fields, as well as the Green's function and the
transmission of such systems, has been developed and demonstrated.\cite{Code} It includes estimation of the
accuracy and convergency of calculations and in particular extrapolation of the eigen-wavevectors
towards their exact values which are generally not available. Particular
examples which illustrate the general method and the developed algorithm include a dielectric slab
with wide-layer and $\delta$-perturbations as well as an optical microcavity having
different number of Bragg mirrors. In these examples, a comparison with exact solutions has been
made in order to verify the approach. In all three systems the resonant states and
the transmission are reproducible to any required accuracy by the resonant state expansion. The
wave vectors of resonant states are the most essential part of the calculation as they most
strongly affect the optical properties of the system through the poles of the Green's function. The
extrapolation of the wave vectors using the power law in the basis size, which has been developed
and demonstrated, significantly improve the accuracy of calculations, by one or two orders of
magnitude. Application of the method to two and three-dimensional systems will be reported in future works.

\acknowledgments M.\,D. acknowledges support of EPSRC under the DTA scheme.

\appendix
\section{Analytic calculation of resonant states and perturbation matrices}
\label{App1}

\subsection{Resonant states of the unperturbed slab}

Solving the wave equation Eq.\,(\ref{pslab}) with $\Delta\varepsilon(z)=0$ and the profile of the
dielectric constant $\varepsilon(z)$ given by Eq.\,(\ref{unpslab}), the electric field of RS $n$,
normalized according to Eq.\,(\ref{nint}), takes the form

\begin{equation}
E_n(z)=\left\{
\begin{array}{lll}
(-1)^nA_ne^{-ik_nz}\,, & & z<-a\,,\\
B_n[e^{i\sqrt{\epsilon_s}k_nz}+(-1)^ne^{-i\sqrt{\epsilon_s}k_nz}]\,, &  &|z|\leq a\,,\\
A_ne^{ik_nz}\,, && z>a\,,
\end{array} \right.
\label{basise1}
\end{equation}
where
\begin{equation}
A_n=\frac{e^{-ik_na}}{\sqrt{a(\epsilon_s-1)}}\,,\qquad B_n=\frac{(-i)^n}{2\sqrt{a\epsilon_s}}\,.
\label{basise2}
\end{equation}
The RS wave vectors are given by
\begin{equation}
k_n=\frac{1}{2a\sqrt{\epsilon_s}}(\pi n-i\ln\gamma),\qquad n=0,\,\pm1,\,\pm2,\,...\,, \label{basise3}
\end{equation} with
\begin{equation}
\gamma=\frac{\sqrt{\epsilon_s}+1}{\sqrt{\epsilon_s}-1}\,, \label{basise4}
\end{equation}
all having the same imaginary part.

\subsection{Resonant states of a slab perturbed by a wide dielectric layer}

The exact solutions of the wave equation Eq.\,(\ref{pslab}) for the system with the perturbation
given by Eq.\,(\ref{pone}) and outgoing boundary conditions has the form
\begin{equation}
\mathcal{E}^{\rm{(exact)}}_{\nu} (z)=\left\{
\begin{array}{ll}
A_{\nu}e^{-i\varkappa_{\nu}z}\,, & z<- a\,,\\
B_{\nu} e^{i\sqrt{\epsilon_s}\varkappa_{\nu}z}+C_{\nu}e^{-i\sqrt{\epsilon_s}\varkappa_{\nu}z}\,, & {-a}\leq z \leq b\,, \\
D_{\nu} e^{i\sqrt{\epsilon_p}\varkappa_{\nu}z}+E_{\nu}e^{-i\sqrt{\epsilon_p}\varkappa_{\nu}z}\,, &b\leq z \leq a\,, \\
H_{\nu}e^{i\varkappa_{\nu}z}\,, & z>a\,, \\
\end{array} \right.\label{wpl}
\end{equation}
where $\epsilon_p=\epsilon_s+\Delta\epsilon$, and $b=a/2$. The coefficients in Eq.\,(\ref{wpl}) are found
from the continuity of the electric field and its derivative and the normalization condition
Eq.\,(\ref{nint2}). The complex-valued RS wave numbers $\varkappa_{\nu}$ are found by solving a
secular equation following from the boundary conditions:
\begin{equation}
    \beta \gamma f(k)g(k)-1=\frac{\beta-\gamma}{\beta\gamma-1}\Bigl[\beta g(k)-\gamma f(k) \Bigr]\,,
\label{wvnum1}
\end{equation}
where
\begin{equation}
    \beta=\frac{\sqrt{\epsilon_p}+1}{\sqrt{\epsilon_p}-1}\,,
\end{equation}
and the functions $f(k)$ and $g(k)$ are defined as
\begin{equation}
f(k)=e^{-2i\sqrt{\epsilon_s}k(a+b)}\,,\ \ \ g(k)=e^{-2i\sqrt{\epsilon_s}k(a-b)}\,.
\end{equation}
We solve Eq.\,(\ref{wvnum1}) using the Newton-Raphson method to find
$k=\varkappa_{\nu}^{\rm{(exact)}}$.

\subsection{Matrix elements of the wide-layer perturbation}
Using Eq.\,(\ref{pmatrix}) and basis functions Eq.\,(\ref{basise1}) we calculate $V_{nm}$ for the
wide-layer perturbation Eq.\,(\ref{pone}) to be
\begin{eqnarray}\label{Vnm1}
V_{nm}=\frac{\Delta\epsilon}{\epsilon_s}\,\frac{1}{4ia\sqrt{\epsilon_s}}\Bigl[\!\!\!\!\!\!&&(-i)^{n+m}\eta(k_n+k_m,z)\\
&&+(-i)^{n-m}\eta(k_n-k_m,z)\nonumber \\
&&+(-i)^{-n+m}\eta(-k_n+k_m,z)\nonumber \\
&&+(-i)^{-n-m}\eta(-k_n-k_m,z)\Bigr]_b^a\,,\nonumber
\end{eqnarray}
for $n\neq m$ and
\begin{equation}
V_{nn}=\frac{\Delta\epsilon}{\epsilon_s}\,\left\{\frac{a-b}{2a}
+(-1)^n\frac{\bigl[\eta(2k_n,z)+\eta(-2k_n,z)\bigr]_b^a}{4ia\sqrt{\epsilon_s}} \right\}
\end{equation}
for $n=m$, where $\eta(k,z)=e^{i\sqrt{\epsilon_s}kz}/k$.

\subsection{Resonant states of a slab perturbed by a delta scatterer} 

In the case of a $\delta$-perturbation $\Delta\varepsilon(z)=w\epsilon_d\delta(z-b)$ with $|b|\leq a$,
the secular equation for the RS wave vectors takes the form
\begin{equation}\label{T2}
\bigl[1+\gamma f(k)\bigr]\bigl[1+\gamma g(k)\bigr]=\frac{2i\sqrt{\epsilon_s}}{w\epsilon_d k}\bigl[1-\gamma^2 f(k)g(k)\bigr]\,.
\end{equation}
It is also solved numerically with the help of the Newton-Raphson method to find
$k=\varkappa_{\nu}^{\rm{(exact)}}$.

\subsection{Matrix elements of the $\delta$-perturbation}
Using Eq.\,(\ref{pmatrix}) and basis functions Eq.\,(\ref{basise1}) we calculate $V_{nm}$ for the
$\delta$-perturbation to be
\begin{eqnarray}
V_{nm}=w\epsilon_d E_n(a/2)E_m(a/2)\,. \label{Vnm2}
\end{eqnarray}

\end{document}